\documentclass[
twocolumn,
letterpaper,
showpacs,
floatfix, 
aps,prb,amsmath,amssymb,
groupaddress,eqsecnum]{revtex4}
\usepackage{graphicx}
\usepackage{amscd}

\newcommand{\be}{\begin{equation}}
\newcommand{\ee}{\end{equation}}
\newcommand{\bea}{\begin{eqnarray}}
\newcommand{\eea}{\end{eqnarray}}

\newcommand*{\bs}{\begin{split}}
\newcommand*{\bes}{\begin{equation}\begin{split}}
\newcommand*{\ees}{\end{split} \end{equation}}

\begin{document}

\title{Quantum optimal control theory and dynamic \\
coupling in the spin-boson model}

\author{H. Jirari}
\email{hamza.jirari@uni-graz.at}
\author{W.~P\"{o}tz}
\email{walter.poetz@uni-graz.at}
\affiliation{
Institut f\"{u}r Physik, Theory Division
Karl-Franzens-Universit\"{a}t Graz,\\
Universit\"{a}tplatz 5, 8010 Graz, Austria}


\begin{abstract}
A Markovian master equation
describing the evolution of open quantum systems in the presence of 
a time-dependent external field is derived within the Bloch-Redfield formalism. 
It leads to a system--bath interaction which depends on the control
field. Optimal control theory is used to select 
control fields which allow accelerated or decelerated system relaxation,
or suppression of relaxation (dissipation) altogether, depending
on the dynamics we impose on the quantum system.
The control--dissipation correlation and the non-perturbative treatment of the
control field are essential for reaching this goal. The optimal control
problem is formulated within Pontryagin's minimum principle and the resulting optimal
differential system is solved numerically. As an application,
we study the dynamics of a spin-boson model in the strong coupling regime
under the influence of an external control field.
We show how trapping the system in unstable quantum states and transfer of population
can be achieved by optimized control of the dissipative quantum system.
We also used optimal control theory to find the driving 
field that generates the quantum Z-gate. In several cases studied,
we find that the selected optimal field which reduces 
the purity loss significatively is a multi--component low--frequency 
field including higher harmonics, all of which lie below the phonon cutoff frequency. 
Finally,
in the undriven case we present an analytic result for the Lamb shift at zero temperature.
\end{abstract}
\date{\today}
\maketitle

\section{INTRODUCTION\label{I}}

In the theory of quantum information and computation, quantum coherence and entanglement
are used as essential resources for efficient information processing~\cite{Nielson}.
However, the interaction of the quantum system with its environment eventually leads
to a complete loss of the information initially stored in its quantum state.
This phenomena, known as decoherence, is regarded as a serious obstacle
to a successful implementation of quantum information processing.
The question of how it is possible to avoid the negative influence
of this process is one of the most interesting issues in modern quantum mechanics.
It not only concerns the area of quantum information and computation
but many other fields of physics as well. 
The challenge is to preserve quantum coherence 
during a sufficiently long time needed for both storage and manipulation of the quantum 
states in systems which are unavoidably exposed to the influence
of their surrounding environment.

Over the last few years, a number of interesting schemes have been 
proposed to eliminate the undesirable effects of decoherence
in open quantum systems, including decoherence free subspaces~\cite{Zanardi_1,Lidar}, quantum
error correction codes~\cite{Nielson,Preskill,Knill}, quantum feedback
\cite{Wang} and mechanisms based on the unitary ``bang-bang''
pulses and their generalization, quantum dynamical decoupling
\cite{Viola_1,Viola_2,Zanardi_2,Vitali,Uchiyama,Byrd,Protopopescu}.
The key ingredient of dynamical decoupling
is the continuous disturbance of the system, which suppresses the system-environment
interaction. It has been shown that, in the bang-bang control schemes,
the decoherence of the system is effectively suppressed if the pulse
rate is much higher than the decoherence rate due to the system-environment 
interaction. As already pointed out by Viola and Lloyd
\cite{Viola_1,Viola_2}, the situation
is similar to the so-called quantum Zeno effect~\cite{Wayne} which takes place in 
a system subject to frequent measurements projecting it onto
its initial state: if the time interval between two projections is small
enough the evolution of the system is nearly ``frozen''.
In a similar manner to the quantum Zeno effect, a fast rate control 
freezes decoherence. Analysis and comparison of the effects of 
these different physical procedures (bang-bang dynamic decoupling,
coherence protection by the quantum Zeno effect and continuous coupling) 
have been investigated in Ref.~\cite{Facchi}.
Advances in decoherence control using dynamical decoupling startegies 
is addressed in Ref.~\cite{Viola_3}.

The staring point of the decoupling techniques is the observation that even
though one does not have access to the large number of uncontrollable degrees
of freedom of the environment, it is still possible to interfere with its dynamics
by inducing motions into the system which are at least as fast as the environment
dynamics~\cite{Vitali}. Moreover, if one can establish an additional coupling to the system   
by means of an external control, there can be quantum interference 
between the two interactions.  
The degree and nature of quantum interference -- constructive or destructive -- 
can be controlled by adjustment of the control field~\cite{CCP}.

In a simple--minded model of a dissipative quantum system, 
where the interference between the system--bath and system--control 
interaction is ignored or is irrelevant only limited control can be achieved~\cite{Jirari}.
The situation changes dramatically when 
interference between the system--environment and system--control 
interaction can be used to control the effective system--environment
coupling~\cite{CCP,Xu,Kocharovskaya,Schirr,Morillon_1,Morillon_2,Dakhnovskii,Grifoni,Stigmunt_1,Stigmunt_2,Jirari_2}.   
This effect of coherent control of ``dissipation" is demonstrated here for the 
example of the driven spin--boson model in which a quantum two-level system (qubit)
is modelled by a spin, the environmental heat bath by quantum oscillators, 
and the spin subjected to an external control field is coupled
to each bath oscillator independently. Decoherence control 
of this model is formulated using optimal control 
which is mathematically a problem of functional optimization
under dynamical constraints~\cite{Pontryagin,ARTHURE,Betts}. 
Recently, we studied the same model with a control field 
restricted to a monochromatic wave plane~\cite{Jirari_2}.
The task was to find a set of three parameters namely
the amplitude, the frequency and the phase using optimal control theory.
Results were presented for control of the relative 
population of the spin system, {\it i.e}, the z-component of the Bloch vector.
In the present paper, we show how this work can be extended to a control of 
all components of the Bloch vector simultaneously, 
as well as to general control field shapes. 

The spin-boson model is a widely used model system. 
It can be mapped to a number of physical situations~\cite{WEISS}.
In the theory of open quantum systems, the spin-boson model
is actually one of the most popular models and has 
gained recent practical importance in the field 
of quantum computation~\cite{Nielson}. A special variant of it, 
in which the inter--level coupling is absent, is known as the independent--boson model~\cite{Mahan}.   
These models have been used to study the role of the electron--phonon interaction 
in point defects and quantum dots, interacting many--body systems, magnetic
molecules, bath assisted cooling of spins and a two level Josephson-Junction
~\cite{Kuhn,Hohenester,Zhang,Luban,Allahverdyan_1,Shon}. 
Its basic properties have been reviewed in the literature~\cite{Legget}.

The remainder of the paper is organized as follows: in the next section we present
a pedagogical derivation of Born-Markov master equations for dissipative N-level   
systems in the presence of time-dependent external control fields. The master equation
is written as a set of Bloch-Redfield equations and a qualitative discussion of the influence
of the control field on dissipation is given.
This equation is the starting point for the derivation 
of the kinetic equation for the driven spin-boson model in the strong spin--boson coupling
regime, as outlined in Sec.~\ref{III}. 
In Sec.~\ref{IV}, within the Pontryagin minimum principle 
we formulate the optimum control problem 
in terms of the Bloch vector.
The general cost functional and its gradient in case of arbitrary control field are given.
We also present the numerical approach in form of the gradient method.
Our numerical results presented in Sec.~\ref{V}. 
Summary and conclusions are given in
Sec.~\ref{VI}. Some mathematical details are relegated to appendixes.

\section{QUANTUM MASTER EQUATION FOR DRIVEN OPEN SYSTEMS\label{II}}

Consider a physical system $S$ embedded in a dissipative environment 
($B$ also referred to as the heat bath)
and interacting with a time-dependent classical external field, {\it i.e.}, the control field.
The Hilbert space of the total system ${\cal H}_{\rm tot}={\cal H}_S\otimes{\cal H}_B$ is expressed as 
the tensor product of the system Hilbert space ${\cal H}_S$ and the environment Hilbert space ${\cal H}_B$. 
The total Hamiltonian has the general form
\be
H_{\rm tot} =H_S(t)+ H_B+ H_{\rm int}\, ,
\label{Hamiltonian1}
\ee
where $H_S(t)$ is the Hamiltonian of the system, $H_B$ of the bath and 
$H_{\rm int}$ of their interaction that is responsible for decoherence. 
The operators $H_S(t)$ and $H_B$ act on ${\cal H}_S$ and  ${\cal H}_B$, respectively.
The operator $H_S(t)$ contains a time-dependent external field to control
the quantum evolution of the system.

\subsection{Perturbation theory in the system-bath coupling\label{II-A}}

We shall be interested in the time evolution of the reduced density matrix of an open system,
defined as 
\be
\rho_S(t) = {\rm tr}_B\left\{\rho_{\rm tot}(t)\right\}\, .
\ee
where $\rho_{\rm tot}$ is the total density matrix for both the system
and the bath. Here and in the following ${\rm tr}_S$ denotes the partial trace
over the open system's Hilbert space, while ${\rm tr}_B$ denotes the partial trace
over the degrees of freedom of the environment $B$.
A number of different methods are used
to derive an equation of motion for the reduced density matrix~\cite{Breuer}.
However, in any practical applications, two approximations are commonly 
invoked. The first is the initial factorization ansatz. Basically, one
assumes that the interaction $H_{\rm int}$ is switched on at time $t=0$. 
Prior to this, the system $S$ and the environment $B$ 
are uncorrelated so that the initial density matrix 
of the total system factorizes as the product of the system and 
bath contributions, that is, 
\be
\label{eq: INITIAL_FAC}
\rho_{\rm tot}(0)=\rho_S(0)\otimes\rho_{B}
\ee
with $\rho_B = {\rm tr}_S\left\{\rho_{\rm tot}(0)\right\}$
and $\rho_S(0)= {\rm tr}_B\left\{\rho_{\rm tot}(0)\right\}$.
The second approximation is the weak system-bath interaction limit in which the second-order
perturbation theory is applicable. In the following, we shall present a derivation
of the master equation for $\rho_S$. $H_{\rm int}$ is assumed to be weak and will be 
treated perturbatively up to second order for its effects on 
$\rho_S$ under coherent external driving.

Let us start with the equation of motion for the total density matrix, 
\be
\label{eq:VN_SCHR_PIC}
\dot\rho_{\rm tot}(t) = -\frac{i}{\hbar}\left\lbrack H_{\rm tot}(t),\rho_{\rm tot}(t)\right\rbrack \ .
\ee
The formal solution of the above equation 
can be obtained in the interaction picture~\cite{Morillon_1} with respect to the Hamiltonian
$H(t)=H_S(t)+ H_B$,
\be
\label{eq:RHO_INT_PIC}
\rho^{I}_{\rm tot}(t) = U^\dagger(t,0)\rho_{\rm tot}(t)U(t,0) \ , 
\ee
where the unitary operator reads
\be
\label{eq:SCHR_OPERATOR}
U(t,t')={\cal T}\left\{\exp\left\lbrack -{i\over\hbar}\int_{t'}^t\,d\tau\, H(\tau)\right\rbrack\right\} \ .
\ee
The time ordering ${\cal T}$ of the exponential in Eq. (\ref{eq:SCHR_OPERATOR}) 
is defined as the Taylor series with each term being time ordered. 
Using the fact that the operators $H_S(t)$ and $H_B$ commute 
\be
\left\lbrack H_S(t), H_B\right\rbrack = 0
\ee
$U(t,t')$ can be decomposed in the form
\be
\label{eq:DECOMP}
U(t,t') = U_S(t,t')\otimes U_B(t-t')
\ee
with
\be
\label{eq:PROPAGATOR_SYS}
U_S(t,t') = {\cal T}\left\{\exp\left\lbrack -{i\over\hbar}\int_{t'}^t\,d\tau\,H_S(\tau)\right\rbrack\right\} 
\ee
being the propagator of the coherent system dynamics satisfying the Schr\"odinger equation 
\be
\label{eq:SCHRODINGER_EQUATION}
i\hbar\frac{\partial}{\partial t} U_S(t,t')= H_S(t)U_S(t,t') 
\ee
subject to the initial condition 
\be
U_S(t',t')= {\cal I}\, ,
\ee
and 
\be
\label{eq:PROPAGATOR_ENV}
U_B(t-t')=\exp\left\lbrack-\frac{i}{\hbar}H_B(t-t')\right\rbrack
\ee
is the propagator describing the free evolution of the environment.
The equation of motion of $\rho^I_{\rm tot}$ is then
\be
\label{eq:VN_INT_PIC}
\dot\rho^I_{\rm tot}(t)= - \frac{i}{\hbar}\left\lbrack\ H_{I}(t),\rho^I_{\rm tot}(t)\right\rbrack
\ee
with $H_{I}(t)$ defined as 
\be
\label{eq:HAM_INT_PIC}
H_{I}(t)=U^\dagger(t,0) H_{\rm int }U(t,0)\, .
\ee
$H_{I}(t)$ is the interaction Hamiltonian written in the interaction picture.
In integral form, Eq. (\ref{eq:VN_INT_PIC}) can be rewritten  as
\be
\label{eq:VN_INT_PIC_IF}
\rho^I_{\rm tot}(t) = \rho_{\rm tot}(0) - \frac{i}{\hbar}\int_0^{t}\,dt'\left\lbrack\ H_{I}(t'),\rho^I_{\rm tot}(t')\right\rbrack
\ee
Inserting Eq. (\ref{eq:VN_INT_PIC_IF}) into (\ref{eq:VN_INT_PIC}) and taking the trace 
over the bath we find the equation of motion for the reduced density 
matrix of the physical system
\be
\label{eq:VN_INT_PIC_ITER}
\dot\rho_S^I(t)= -\frac{1}{\hbar^2}\int_0^{t}\,dt'{\rm tr}_B\left\{
\left\lbrack H_{I}(t),\left\lbrack H_{I}(t'),\rho^I_{\rm tot}(t')\right\rbrack\right\rbrack\right\}\, . 
\ee
We have assumed that (using Eq.~(\ref{eq: INITIAL_FAC}))
\be
\label{eq:_<Hint_0_>_}
{\rm tr}_B \left\{\left\lbrack H_{I}(t),\rho_{\rm tot}(0)\right\rbrack\right\} =
{\rm tr}_B \left\{\left\lbrack H_{I}(t),\rho_S(0)\otimes\rho_B\right\rbrack\right\}=0
\ee
This assumption has to be justified when considering a specific case.   Equation (\ref{eq:VN_INT_PIC_ITER}) 
still contains the density matrix of the total system
$\rho^I_{\rm tot}(t)$ on its right-hand side. In order to eliminate $\rho^I_{\rm tot}(t)$
from the equation of motion, we assume that the interaction between the system and the environment 
is weak and perform the Born approximation~\cite{Morillon_1,Breuer,Blum} 
\be
\label{eq:BORNAPPROX}
\rho^I_{\rm tot}(t') = \rho^I_S(t')\otimes\rho_B
\ee
In practice, one usually assumes a thermal equilibrium for the environment,
\be
\rho_B= \frac{e^{-\beta H_B}}{{\rm tr}\,e^{-\beta H_B}}\, ,
\ee
where $\beta = 1/k_BT$ with $T$ the environment equilibrium temperature.
Inserting the tensor product (\ref{eq:BORNAPPROX}) in the exact equation of motion
(\ref{eq:VN_INT_PIC_ITER}), we obtain a closed integro-differential equation for the reduced 
density matrix $\rho_S(t)$
\be
\label{eq:VN_INT_PIC_ID}
\dot\rho_S^I(t)= -\frac{1}{\hbar^2}\int_0^{t}\,dt'{\rm tr}_B\left\{
\left\lbrack H_{I}(t),\left\lbrack H_{I}(t'),\rho^I_S(t')\otimes\rho_B\right\rbrack\right\rbrack\right\}\, .
\ee
In order to further simplify the above equation we perform 
the Markov approximation~\cite{Morillon_1,Breuer,Blum}  
in which the integrant $\rho^I_S(t')$ 
is replaced by $\rho^I_S(t)$
\be
\label{eq:VN_INT_PIC_MARKOVAPPROX}
\dot\rho_S^I(t)= -\frac{1}{\hbar^2}\int_0^{t}\,dt'{\rm tr}_B\left\{
\left\lbrack H_{I}(t),\left\lbrack H_{I}(t'),\rho^I_S(t)\otimes\rho_B\right\rbrack\right\rbrack\right\}\, .
\ee
The Markov approximation is therefore justified if the time scale $\tau_R$ over which the state
of the system varies appreciably is large compared to the time scale $\tau_B$ over which
the reservoir correlation functions decay ($\tau_R \gg \tau_B$).
One now goes back to the Schr\"odinger picture using Eqs. 
(\ref{eq:RHO_INT_PIC}), (\ref{eq:HAM_INT_PIC}),
the decomposition (\ref{eq:DECOMP}), and 
the cyclic invariance property of the trace
%
%
\bea
&&\label{eq:VN_SCHROD_PIC}
{\dot\rho}_S(t)=-\frac{i}{\hbar}[H_S(t),\rho_S(t)]\nonumber\\
&&~~-\frac{1}{\hbar^2}\int_0^{t}\,dt'\mbox{tr}_B\left\{\left\lbrack H_{\rm int}(t-t'),
\left\lbrack \tilde{H}_{\rm int}(t,t'),\rho_S(t)\otimes\rho_B\right\rbrack\right\rbrack\right\}\,.
\nonumber\\
\eea
%
%
Here
\be
\tilde{H}_{\rm int}(t,t')=U_S(t,t')H_{\rm int}U_S^\dagger(t,t')
\ee
is the interaction Hamiltonian in the anti-Heisenberg representation with respect to $H_S(t)$ and 
\be
{H}_{\rm int}(t-t')=U_B^\dagger(t-t')H_{\rm int}U_B(t-t')
\ee 
the interaction Hamiltonian in the Heisenberg representation
with respect to $H_B$. The time-evolution operators $U_S(t,t')$ and $U_B(t-t')$ are defined by 
Eqs.~(\ref{eq:PROPAGATOR_SYS}) and (\ref{eq:PROPAGATOR_ENV}), respectively.

\subsection{Bloch-Redfield equations\label{II-B}}

Let us suppose that ${\cal H}_S$ is an $N$-dimensional Hilbert space of orthonormal basis 
$| i \rangle, i=1\ldots N$. Writing the quantum master equation (\ref{eq:VN_SCHROD_PIC}) in the representation 
$| i \rangle$ and expanding the double commutator, we obtain after some algebra~\cite{Morillon_1,Blum}
%
%
\bea
\label{eq:MASTER_EQUATION}
{\dot\rho}_{S,ij}(t) &=&-\frac{i}{\hbar}\sum_{kl}\left(H_{S,ik}(t)\delta_{lj}
-\delta_{ik}H_{S,lj}(t)\right)\,{\rho}_{S,kl}(t)
\nonumber\\
&&\qquad-\sum_{kl}{\cal R}_{ijkl}(t)\,{\rho}_{S,kl}(t)
\eea
%
where the first term represents the unitary part of the dynamics
generated by the system Hamiltonian $H_S(t)$ and the second term 
accounts for the dissipative effects due to the 
coupling of the system to the environment.
The Redfield relaxation tensor $R_{ijkl}(t)$  is given by 
\bea
\label{eq:REDFIELD_TENSOR}
{\cal R}_{ijkl}(t)&=&\delta_{lj}\sum_r\,\Gamma_{irrk}^+(t)
+\delta_{ik}\sum_r\,\Gamma_{lrrj}^-(t)\nonumber\\
&&\qquad-\Gamma_{ljik}^+(t) -\Gamma_{ljik}^-(t)
\eea
with the time-dependent rates $\Gamma_{ijkl}^{\pm}(t)$ 
%
%
\begin{subequations}
\bea
\label{eq:Gamma+}
\Gamma_{ljik}^+(t)&=&\frac{1}{\hbar^2}
\int_0^{t}\,dt'\,\mbox{tr}_B\left\{ 
H_{{\rm int},lj}(t-t'){\tilde H}_{{\rm int},{ik}}(t,t')
\rho_B\right\}\, ,
\nonumber\\
&&\\
\label{eq:Gamma-}
\Gamma_{ljik}^-(t)&=&\frac{1}{\hbar^2}
\int_0^{t}\,dt'\,\mbox{tr}_B\left\{
{\tilde H}_{{\rm int},{lj}}(t,t')H_{{\rm int},{ik}}(t-t')\rho_B\right\}\, .
\nonumber\\
\eea
\end{subequations}
Let us now write the Hamiltonian $H_{\rm int}$ in the Schr\"odinger picture 
in the following general form
\be
\label{eq:HAM_SCHRO_PIC}
H_{\rm int}= \sum_{\alpha}\, A_\alpha\otimes B_\alpha
\ee 
where $A_\alpha$ and $B_\alpha$ are hermitian operators of the system and the environment,
respectively. Inserting the expression (\ref{eq:HAM_SCHRO_PIC}) into Eqs. (\ref{eq:Gamma+}) and (\ref{eq:Gamma-}),
using the fact that $A_\alpha$ and $B_\alpha$ commute leads to~\cite{Morillon_1,Blum}

\begin{subequations}
\bea
\label{eq:Gamma+_BIS}
&&\Gamma_{lj,ik}^+(t)=\frac{1}{\hbar^2}\int_0^{t}\,dt'
\sum_{\alpha,\beta}\left\langle B_\alpha(t-t')B_\beta(0)\right\rangle_B
A_{\alpha,{lj}}\nonumber\\
&&\times\sum_{m,n} U_{S,{im}}(t,t')A_{\beta,{mn}} U_{S,{kn}}^{\ast}(t,t')\,,
\nonumber\\
&&\\
\label{eq:Gamma-_BIS}
&&\Gamma_{lj,ik}^-(t)=\frac{1}{\hbar^2}\int_0^{t}\,dt'
\sum_{\alpha,\beta}\left\langle B_\beta(0)B_\alpha(t-t')\right\rangle_B 
\nonumber\\
&&\times\sum_{m,n} U_{S,{lm}}(t,t')A_{\beta,{mn}}U_{S,{jn}}^{\ast}(t,t')\,A_{\alpha,{ik}}\,,
\nonumber\\
\eea
\end{subequations}

where
\be
\left\langle B_\alpha(\tau)B_\beta(0)\right\rangle_B={\rm tr}_B\left\{ B_\alpha(\tau)B_\beta(0)\rho_B\right\}
\ee
is the environment correlation function with 
$\left\langle B_\alpha(0)B_\beta(\tau)\right\rangle_B = \left\langle B_\alpha(-\tau)B_\beta(0)\right\rangle_B$,
since the environment is supposed to be in thermal equilibrium and its evolution is then homogeneous in time.
Note that the condition (\ref{eq:_<Hint_0_>_}) becomes
\be
\label{eq:first_moment}
\left\langle B_\alpha(\tau)\right\rangle_B={\rm tr}_B\left\{ B_\alpha(\tau)\rho_B\right\}=0,
\ee
which states that the reservoir averages of $B_\alpha(\tau)$ vanish.

The Born-Markov master equation (\ref{eq:MASTER_EQUATION}) with the time-dependent decay rates 
defined in Eqs. (\ref{eq:Gamma+_BIS}) and (\ref{eq:Gamma-_BIS})
together with the Schr\"odinger equation (\ref{eq:SCHRODINGER_EQUATION})
satisfied by the coherent time evolution operator $U_S(t,t')$
provide all the necessary ingredients to describe the dynamics 
of a driven open quantum system. Note that the interaction of the system 
with the time-dependent control Hamiltonian $H_S(t)$ is treated 
non-perturbatively in the derivation of the
above quantum master equation. The application of the present formulation 
to the driven spin--boson model in the weak coupling regime enables us to find an identical
set of Bloch-Redfield equations obtained from projector-operator
methods by Hartmann {\it et al.}~\cite{Hartmann}. 

In the absence of the control field, $U_S(t,t')\equiv U_S(t-t')$
is the free time evolution operator of the physical system. In this particular case,
the time integration in Eqs. (\ref{eq:Gamma+_BIS}) and (\ref{eq:Gamma-_BIS}) can be replaced by 
$\frac{1}{2}\int_{-\infty}^{+\infty}$ in the Markov approximation. 
After the extention of time integration to infinity, 
the decay rates $\Gamma_{ijkl}^{\pm}$ become time-independent and are 
then given in terms of the Fourier transform of the product of the environment 
correlation functions and the system dissipative operators
(the interaction vertices), in agreement with the well-known results
for undriven quantum open systems~\cite{Blum}.

\subsection{Control field effects\label{II-C}}

The analysis of Eq. (\ref{eq:MASTER_EQUATION})
shows that the presence of a time-dependent external field 
affects both the unitary evolution
and the dissipative parts of the quantum master equation.
The dissipative field influence can be interpreted as 
a direct consequence of quantum interference between the system-bath interaction
and a coupling of the system to the external field. In order to  
study the dissipative field influence, let us examine 
the transition and the dephasing rates in the secular approximation~\cite{Breuer,Blum}.
In the zero field limit, 
the secular approximation assumes that the populations and the coherences are 
completely decoupled. It is valid if the typical time scale of the evolution 
of the system $\tau_S$, defined by a typical value for the inverse of 
the frequency associated with the system's energy levels,  
is small compared to the relaxation time of the system, that is $\tau_S\ll\tau_R$. 
If we suppose that the secular approximation is still valid in the non-zero field case,
the transition rates are defined as~\cite{Kocharovskaya,Schirr,Blum}
\bea
W_{ij}(t)&=&{\cal R}_{ii,jj}(t)\nonumber\\
&=&\Gamma_{ji,ij}^+(t) + \Gamma_{ji,ij}^-(t)\,,
\eea
with $i \not= j$. Then, from the property 
$
\left\lbrack\Gamma_{lj,ik}^-(t)\right\rbrack^*=\Gamma_{ki,jl}^+(t),
$
it follows that the time dependent parameters $W_{ij}(t)$ are real,
$
W_{ij}^*(t)=W_{ij}(t)
$.
On the other hand, the dephasing rates are defined as~\cite{Kocharovskaya,Schirr,Blum}  
\bea
&&\Gamma_{ij}(t)={\cal R}_{ij,ij}(t)=
\sum_r\left\lbrack\Gamma_{ir,ri}^+(t) + \Gamma_{jr,rj}^-(t)\right\rbrack
\nonumber\\
&&\qquad\qquad-\Gamma_{jj,ii}^+(t)- \Gamma_{ii,jj}^-(t)\, .
\eea
The hermiticity condition of the density matrix implies
$
\Gamma_{ij}^*(t)=\Gamma_{ji}(t)
$
which means that ${\Gamma}_{ij}(t)$ are complex numbers. 
In the zero field limit, the parameters $W_{ij}$ lead
to a population relaxation into a thermal mixture of the system's energy
eigenvalues. Therefore, the diagonal elements of the density
matrix decay to the value given by the Boltzmann factors.
The real part of the parameters $\Gamma_{ij}(t)$ describes
the dephasing, namely the decay of the off-diagonal elements of the density matrix
towards zero. Their imaginary part leads
to a renormalization of the system Hamiltonian which is induced by the vacuum
fluctuation of the environment quantum operators (Lamb Shift).
If a time-dependent external control field is applied, all these quantities 
become time-dependent (via the control field), and an external control
of the dissipation is then possible. 
In particular, the correlation between 
the control field and the dissipation 
leads to the destruction of the detailed balance
\be
\label{eq:detailed_balance}
\lim_{t\to\infty}{\frac{W_{ij}(t)}{W_{ji}(t)}} \not= \frac{\exp(-\beta E_i)}{\exp(-\beta E_j)}\, .
\ee
and allows for the control of the states populations. Here $E_i$ are the energy
eigenvalues of the undriven physical system. Eq. (\ref{eq:detailed_balance})
shows that the steady state can be far from equilibrium in the presence of a 
control field.

Taking the limit $\tau_B\to 0$ as a reasonable approximation, gives
\be
\label{eq:DELTA_CORRE}
\left\langle B_\alpha(\tau)B_\beta(0)\right\rangle_B\propto \delta_{\alpha\beta}\delta(\tau)
\ee
A random interaction with a $\delta$-correlation function is called white-noise,
because the spectral distribution which is given by the Fourier transform
of (\ref{eq:DELTA_CORRE}) is then independent of the frequency~\cite{Risken}.
Substituting Eq. (\ref{eq:DELTA_CORRE}) into Eqs. (\ref{eq:Gamma+_BIS}) and (\ref{eq:Gamma-_BIS}),
any field dependence disappears because $U_S(t,t)={\cal I}$.
For classical problems the white-noise approximation holds when,  
in the high temperature limit,  the environment resides within the classical regime 
and quantum effects may be ignored. In such a situation, the control-dissipation correlation 
disappears.

\section{DRIVEN spin--boson MODEL\label{III}}

\subsection{The model}
The driven spin--boson model consists of a two-level system interacting with
a thermal bath in the presence of a time-dependent external control~\cite{Grifoni,Legget}.
The Hamiltonian for this model can be written as 
\be
\label{eq:spinbosonH}
H_{\rm tot} =H(t)+H_{\rm int}=H_S(t)+ H_B+ H_{\rm int}\, ,
\ee
where the dynamics of the system $S$ is described by the Hamiltonian
\be
\label{eq:spinbosonHS_t}
H_S(t) = -\frac{\hbar}{2}\left(\Delta\sigma_x+\varepsilon_{0z}\sigma_z\right)
-\frac{\hbar}{2}\varepsilon_z(t)\sigma_z\, .
\ee
Here $\sigma _{\alpha}$ with $\alpha=x,y,z$ are the Pauli spin matrices; $\hbar\Delta $ 
is the tunnelling splitting, $\hbar\varepsilon_{0z}$ is an energy bias
and $\hbar\varepsilon_z(t)$ is its modulation by a time-dependent external control
field. The Hamiltonian of the environment is assumed to be composed of
harmonic oscillators with natural frequencies $\omega_{i}$ and masses $m_{i}$,
\be
H_B=\sum_{i=1}^N\left( \frac{p_{i }^{2}}{2m_{i }}+\frac{m_{i}}{2}x_{i }^{2}\omega _{i}^{2}\right)\, ,
\ee
where ($ {x}_1,..., {x}_N, {p}_1,..., {p}_N)$ are the coordinates
and their conjugate momenta.
The interaction between the system S and the environment B is
assumed to be bilinear,
\be
H_{\rm int}=\sum_{i=1}^N\,c_{i}\frac{q_{0}}{2}\sigma _{z}x_{i}\, ,
\ee
where $c_i$ is the coupling constant between the spin coordinate 
and the $i$th environment oscillator with coordinate $q_i$
while $q_{0}$ measures the distance between the left and
right potential wells. The coupling constants enter the spectral
density function $J(\omega)$ of the environment defined by,
\be
\label{eq:discrete_J_omega}
J(\omega)=\frac{\pi}{2}\sum_{i}\,{c_i\over m_i\,\omega_i}\,\delta(\omega-\omega_i)\,.
\ee

\subsection{Polaron transformation\label{III-A}}

The evaluation of the time-dependent Bloch-Redfield tensor ${\mathcal R}_{ijkl}(t)$
for the Hamiltonian (\ref{eq:spinbosonH}), requires knowledge of the propagator 
of the coherent system dynamics $U_S(t,t')$. 
Obtaining an analytical expression for $U_S(t,t')$ is not trivial
because the Hamiltonian of the physical system (\ref{eq:spinbosonHS_t}), 
is time-dependent and not diagonal. 
To get round this difficulty we perform the polaron
transformation of the Hamiltonian (\ref{eq:spinbosonH}).
This transformation is defined by the unitary operator\cite{Legget}
\be
{\cal V}= e^{-\frac{i}{2}\sigma_z\Omega },
\ee
with 
\be
\Omega =\sum_i\Omega_i,\quad\Omega_i= \left(q_0 c_i/\hbar m_i\omega_i^2\right) p_i
\ee
Applied to the original Hamiltonian (\ref{eq:spinbosonH}) leads to
\be
\label{eq:spinbosonH_Trans}
H'_{\rm tot}={\cal V}H_{\rm tot}{\cal V}^{-1}=H'_S(t)+ H'_B+ H'_{\rm int}\,.
\ee
The expression
\be
H'_S(t)=-\frac{\hbar}{2}\left(\varepsilon_{0z}+\varepsilon_z(t)\right)\sigma_z\, ,
\ee
is the internal system part of the transformed Hamiltonian
\be
H'_B=\frac{1}{2}\sum_{i}\left(\frac{p_i^2}{m_i} +m_i\omega_i^2 x_i^2\right)\, ,
\ee
defines the Hamiltonian of the bath, and
\be
\label{eq:spinbosonH_Trans_int} 
H'_{\rm int}=-\frac{1}{2}\hbar\Delta\left(\sigma_+ e^{-i\Omega}+\sigma_- e^{i\Omega}\right)\, ,
\ee
gives the modified interaction. Here 
$\sigma_{\pm}=(\sigma_x\pm i\sigma_y)/2$.
After applying the polaron transformation the coherent propagator
\be
U_S(t,t')={\cal T}\left\{\exp\left\lbrack -{i\over\hbar}\int_{t'}^t\,d\tau\, H'_S(\tau)\right\rbrack\right\}
\ee
simplifies to 
\bea
\label{eq:cohpropa}
&&U_S(t,t')= \cos\left[\left\{\varepsilon_{0z}(t-t')+f(t,t')\right\}/2\right]{\cal I}\nonumber\\
&&\qquad + i\sin\left[\left\{\varepsilon_{0z}(t-t')+f(t,t')\right\}/2\right]\sigma_z
\eea
where ${\cal I}$ is the unit matrix of order $2\times 2$ and  
\be
\label{eq:field_effect_Func}
f(t,t')= \int _{t'}^{t}\,d\tau\varepsilon_z(\tau)\, .
\ee
A constant phonon--induced energy shift in the system Hamiltonian has been omitted.  
As an application of the general formulation of the kinetic equation
for driven open systems developed in Sec. (\ref{II}), we will consider the Hamiltonian 
(\ref{eq:spinbosonH_Trans}) and derive the explicit form of the corresponding master equation
for small $\Delta$. 

The physical situation described by this model can be envisioned as a 
weakly coupled (via $\Delta$) semiconductor double quantum dot containing a single electron.  
Each quantum dot provides one electronic level (spin is ignored).
A metallic gate is located under each quantum dot.  
The relative 
bias between the two electrodes represents the control field. 
Alternatively, the Hamiltonian studied here 
can be interpreted as a spin 1/2--particle in a 
weak static B--field ($\varepsilon_0$, $\Delta$) which is directed slightly 
away from the z--direction in an external control field 
applied in z--direction.   
 
For the transformed Hamiltonian, the interaction contribution is 
not the system--bath interaction, but rather accounts for a phonon--renormalized coupling between the two 
states "up" and "down".  Here, the dissipative operators of the system and those for the environment 
are $S_1=\hbar\Delta\sigma_+/2$, $S_2=S_1^{\dagger}=\hbar\Delta\sigma_-/2$ and $B_1=e^{-i\Omega}$,  
$B_2=B_1^{\dagger}=e^{i\Omega}$, respectively
\footnote{Note that Eq. (\ref{eq:first_moment}) is satisfied by our polaron bath operators
$e^{\pm i\Omega(t)}$. Since $\langle e^{\pm i\Omega(t)}\rangle_B = \langle e^{\pm i\Omega(0)}\rangle_B =e^{-\Phi}$
with $\Phi=\frac{q_0^2}{\pi\hbar}\int_0^{\infty}\,d\omega \frac{J(\omega)}{\omega^2}\coth(\hbar\omega\beta/2)$
which is an infinite integral in case of the Ohmic bath spectral density (\ref{eq:Ohmic_SP_D})
and implies $\langle e^{\pm i\Omega(t)}\rangle_B=0.$}.
For the new interaction Hamiltonian, Eq.~(\ref{eq:spinbosonH_Trans_int}), 
the rates $\Gamma_{ij,kl}^{\pm}(t)$ defined by Eqs. 
(\ref{eq:Gamma+_BIS}) and (\ref{eq:Gamma-_BIS}) may be written in terms
of the equilibrium correlation functions with respect to the 
bosonic bath spectral density $J(\omega)$ in Eq.~(\ref{eq:discrete_J_omega}),
\begin{subequations}
\bea
\label{eq:corr_1}
\langle e^{\pm i\Omega(t)}\,e^{\pm i\Omega(0)}\rangle_B &=& e^{-Q_2(t)}e^{iQ_1(t)}\,,\\
\langle e^{\pm i\Omega(t)}\,e^{\mp i\Omega(0)}\rangle_B &=& e^{-Q_2(t)}e^{-iQ_1(t)}\,,\\
\langle e^{\pm i\Omega(0)}\,e^{\pm i\Omega(t)}\rangle_B &=& e^{-Q_2(t)}e^{-iQ_1(t)}\,,\\
\label{eq:corr_4}
\langle e^{\pm i\Omega(0)}\,e^{\mp i\Omega(t)}\rangle_B &=& e^{-Q_2(t)}e^{iQ_1(t)}\,,
\eea
\end{subequations}
where the exponents are given by~\cite{Mahan}
\begin{subequations}
\bea
\label{eq:BIS_Q_1}
Q_1(t)&=&\frac{q_0^2}{\pi\hbar}\int_0^{\infty}\,d\omega \frac{J(\omega)}{\omega^2}\sin(\omega t)\,,\\
\label{eq:BIS_Q_2}
Q_2(t)&=&\frac{q_0^2}{\pi\hbar}\int_0^{\infty}\,d\omega \frac{J(\omega)}{\omega^2}(1-\cos(\omega t))\coth(\hbar\omega\beta/2)\,.
\nonumber\\
\eea
\end{subequations}

\subsection{Ohmic spectrum of the bath\label{III-D}}

Within the present work, we consider the case of Ohmic spectrum
\be
\label{eq:Ohmic_SP_D}
J(\omega)=\eta\omega e^{-\omega/\omega_c}=\left(2\pi\hbar\alpha/q_0^2\right)\omega e^{-\omega/\omega_c}\, .
\ee
Here $\eta$ is a phenomenological friction coefficient while $\alpha$
is the dimensionless coupling constant, and $\omega_c$ is an ultraviolet frequency cutoff.
For the Ohmic bath (\ref{eq:Ohmic_SP_D}),
the functions $Q_1(\tau)$ in Eq.~(\ref{eq:BIS_Q_1})
and $Q_2(\tau)$ in Eq.~(\ref{eq:BIS_Q_2}) 
can be determined explicitly~\cite{Legget}
\begin{subequations}
\bea
\label{eq:Q_1}
Q_1(\tau)&=& 2\alpha\arctan(\omega_c\tau),\\
\label{eq:Bis_Q_2}
Q_2(\tau)&=&2\alpha\ln
\left
\lbrack
\frac{{\bf\Gamma}^2\left(1+{1\over\hbar\omega_c\beta}\right)\sqrt{1+\omega_c^2\tau^2}}
{{\bf\Gamma}\left(1+{1\over\hbar\omega_c\beta}-i{\tau\over\hbar\beta}\right)
{\bf\Gamma}\left(1+{1\over\hbar\omega_c\beta}+i{\tau\over\hbar\beta}\right)}
\right\rbrack\nonumber\\
\eea
\end{subequations}
where ${\bf\Gamma}$ is Euler's Gamma function.

Let us determine the behaviour of the function $Q_2(\tau)$
at low temperature. Using the relation
\be
{\bf\Gamma}\left(1-i{\tau\over\hbar\beta}\right)
{\bf\Gamma}\left(1+i{\tau\over\hbar\beta}\right)=
\frac{\pi\tau}{\beta\hbar}\frac{1}{\sinh\left(\pi\tau/\beta\hbar\right)}
\ee
one obtains from Eq.~(\ref{eq:Bis_Q_2}) for $\hbar\omega_c\beta\gg 1$
(low temperature)
\be
\label{eq:Q_2}
Q_2(\tau)=\alpha\ln(1+\omega_c^2\tau^2)+
2\alpha\ln\left\lbrack
\frac{\beta\hbar}{\pi\tau}\sinh\left(\frac{\pi\tau}{\beta\hbar}\right)
\right\rbrack\,.
\ee
The first term is independent of temperature and describes
how the fluctuations of the field vacuum affect
the coherence of the open system. This contribution of 
$Q_2(\tau)$ to dissipation depends on the cutoff frequency $\omega_c$.
The second term is the thermal contribution of $Q_2(\tau)$ to dissipation.
Notice its dependence on the thermal correlation time $\tau_B =\beta\hbar/\pi$.
On the other hand, the function $Q_1(\tau)$ is independent of temperature.

\subsection{Master equation}

For the description of the dynamics of a two-level system,
it is convenient to map the state density matrix onto the Bloch vector
${\bf p}(t)=\left(p_x(t),p_y(t),p_z(t)\right)^T\in {\mathbb{R}^3}$
defined by 
\be 
{\bf p}(t)= {\rm Tr}({\mbox{\boldmath $\sigma$}}\rho(t)) 
= \left(\begin{array}{c}
\rho_{01}(t)+\rho_{10}(t)\\
i(\rho_{01}(t)-\rho_{10}(t))\\
\rho_{00}(t)-\rho_{11}(t)
\end{array}\right),
\label{Bloch-vector}
\end{equation}
where ${\mbox{\boldmath $\sigma$}}=(\sigma_x, \sigma_y, \sigma_z)$
is the vector composed of the three Pauli matrices.
Within this notation, the states of a two-level system are parametrized by a 3-component vectors in the Bloch-sphere
$B:=\left\{{\bf p}\in{\mathbb{R}^3};\,\Vert{\bf p}\Vert \le 1\right\}$.

By combining the Redfield equation (\ref{eq:MASTER_EQUATION}) with
Eq.~(\ref{Bloch-vector}), we obtain for the Bloch vector the inhomogeneous linear equation
of motion,
\bea
\label{eq:state_equation}
&&\dot{\bf p}(t) = 
\left(
{\mbox{\boldmath $\varepsilon$}}_0 +{\mbox{\boldmath$\varepsilon$}}(t)\right)\times{\bf p}(t)+
{\mbox{\boldmath $\xi$}}(t)
\times
\left(
{\bf p}(t)\times{\bf n}
\right)\nonumber\\
&&~~~~~~~~~~~~~~~
-{\mbox{\boldmath $\gamma$}}(t)\,{\bf p}(t)+{\mbox{\boldmath $\gamma$}}_0(t)
\eea

\noindent with ${\mbox{\boldmath $\varepsilon$}}_0= (0,0,-\varepsilon_{0z})^T$, 
${\mbox{\boldmath $\varepsilon$}}(t)= (0,0,-\varepsilon_z(t))^T$,
${\bf n} = (0,1,0)^T$, 
\be
{\mbox{\boldmath $\xi$}}(t)=
\left(\begin{array}{c}
\xi(t)\\
0\\
0
\end{array}\right)
\, ,\qquad
{\mbox{\boldmath $\gamma$}}_0(t)=
\left(\begin{array}{c}
0\\
0\\
\gamma_0(t)
\end{array}\right)
\ee
and the relaxation matrix
\be
\label{eq:MG}
{\mbox{\boldmath $\gamma$}}(t)
= \left(\begin{array}{c c c}
      0    & 0& 0 \\
      0& \gamma(t)   & 0 \\
      0            &  0           & \gamma(t)
\end{array}\right).
\ee
where 
\begin{subequations}
\bea
\xi(t)&=&-{\rm Im}\left\lbrack{\mathcal{R}}_{12,12}(t)+{\mathcal{R}}_{12,21}(t)\right\rbrack\nonumber\\
&=&-{\rm Im}\left\lbrack\Gamma^+_{12,21}(t)+\Gamma^-_{21,12}(t)-\Gamma^+_{12,12}(t)-\Gamma^-_{12,12}(t)\right\rbrack\nonumber\\
&&\\
\gamma(t)&=&{\rm Re}\left\lbrack{\mathcal{R}}_{11,11}(t)-{\mathcal{R}}_{11,22}(t)\right\rbrack\nonumber\\
&=&2{\rm Re}\left\lbrack\Gamma^+_{12,21}(t)+\Gamma^+_{21,12}(t)\right\rbrack\\
\mbox{and}\nonumber\\
\gamma_{0}(t)&=&-{\rm Re}\left\lbrack{\mathcal{R}}_{11,11}(t)+{\mathcal{R}}_{11,22}(t)\right\rbrack\nonumber\\
&=&-2{\rm Re}\left\lbrack\Gamma^+_{12,21}(t)-\Gamma^+_{21,12}(t)\right\rbrack
\eea
\end{subequations}
are linear combinations of the Redfield tensor elements.
Eqs.~(\ref{eq:corr_1})-(\ref{eq:corr_4}),
together with Eqs.~(\ref{eq:Gamma+_BIS}) and (\ref{eq:Gamma-_BIS}) for the rates $\Gamma_{ij,kl}^{\pm}(t)$ ,
and the analytical expression of the coherent propagator (\ref{eq:cohpropa}) lead to
\bea
\label{eq:gamma_yx}
&&\xi(t)=\Delta^2\int_0^{t}\,d\tau\,e^{-Q_2(\tau)}\sin[\varepsilon_{0z}\tau +f(t,t-\tau)]\cos[Q_1(\tau)]\nonumber\\
&&\\
\label{eq:gamma_zz}
&&\gamma(t)=\Delta^2\int_0^{t}\,d\tau\,e^{-Q_2(\tau)}\cos[\varepsilon_{0z}\tau+f(t,t-\tau)]\cos[Q_1(\tau)]\nonumber\\
&&\\
\label{eq:gamma_z0}
&&\gamma_{0}(t)=\Delta^2\int_0^{t}\,d\tau\,e^{-Q_2(\tau)}\sin[\varepsilon_{0z}\tau+f(t,t-\tau)]\sin[Q_1(\tau)]\nonumber\\
&&
\eea
where the function $f(t,t')$ is given in Eq.~(\ref{eq:field_effect_Func})
while the functions $Q_1(t)$ and $Q_2(t)$ are defined in Eqs.~(\ref{eq:BIS_Q_1})
and (\ref{eq:BIS_Q_2}), respectively.

The control field dependence of the rates defined in Eqs. (\ref{eq:gamma_yx}), 
(\ref{eq:gamma_zz}) and the inhomogeneous term Eq. (\ref{eq:gamma_z0}) enters via the function $f(t,t')$ 
in Eq. (\ref{eq:field_effect_Func}). The quantity ${\mbox{\boldmath $\xi$}}(t)$ describes  renormalization effects
on the system Hamiltonian since it depends on the imaginary part of the Redfield tensor elements. 
It serves as an effective local--control field correction acting on the system.
The relaxation and dephasing processes are determined by the rate ${\mbox{\boldmath $\gamma$}}(t)$.
Note that the values for ${\mbox{\boldmath $\xi$}}(t)$, ${\mbox{\boldmath $\gamma$}}(t)$
and ${\mbox{\boldmath $\gamma$}}_0(t)$ at the current time $t$
depend on the control field ${\mbox{\boldmath $\varepsilon$}}(t'')$ for $t''\in [0, t]$.

The explicit equations of motion for the components of the Bloch vector reads
\begin{subequations}
\bea
\label{eq:px}
\dot p_x(t) &=& \left(\varepsilon_{0z} + \varepsilon_z(t)\right) p_y(t)\,,\\
&&\nonumber\\
\label{eq:py}
\dot p_y(t)&=& -\left(\varepsilon_{0z} + \varepsilon_z(t)+\xi(t)\right)p_x(t) - \gamma(t)\,p_y(t)\,,
\nonumber\\
&&\\
\label{eq:pz}
\dot p_z(t)&=&-\gamma(t) p_z(t) +\gamma_{0}(t)\, .
\eea
\end{subequations}

We would like to emphasize here that the decoupling of the populations
from the coherences follows from the polaron transformation and the 
perturbative expansion up to second order in the tunnelling coupling 
$\Delta$, and thus no secular approximation is required. 

\subsection{Undriven case}

In order to illustrate the effects of the bath, namely
the relaxation of the system and its energy renormalisation,
we can analyse the master equation in the absence of the control field. 
The analytical expressions for the rates are worked out in the Appendixes.
At zero temperature and for $\alpha > 1/2$, the decay rate follows as
\begin{subequations}
\bea
\left.\gamma(\varepsilon_{0z}> 0)\right|_{T=0}&=&
\frac{\pi\Delta^2}{2{\bf\Gamma}(2\alpha)}\left(\frac{1}{\omega_c}\right)^{2\alpha}
\varepsilon^{2\alpha-1}_{0z}\,e^{-\left(\varepsilon_{0z}/\omega_c\right)}\,,\nonumber\\
&&\\
\left.\gamma(\varepsilon_{0z}< 0)\right|_{T=0}
&=&\frac{\pi\Delta^2}{2{\bf\Gamma}(2\alpha)}\left(\frac{1}{\omega_c}\right)^{2\alpha}
\left(-\varepsilon_{0z}\right)^{2\alpha-1}\,e^{\left(\varepsilon_{0z}/\omega_c\right)}\,,\nonumber\\
&&
\eea
\end{subequations}
which agrees in leading order in $\varepsilon_{0z}/\omega_c$
with the result of Ref.~\cite{Legget}. Here ${\bf\Gamma}$ is the 
Euler's Gamma function. A similar expression holds for the inhomogeneous term 
\begin{subequations}
\bea
\left.\gamma_0(\varepsilon_{0z}> 0)\right|_{T=0}&=&
\frac{\pi\Delta^2}{2{\bf\Gamma}(2\alpha)}\left(\frac{1}{\omega_c}\right)^{2\alpha}
\varepsilon^{2\alpha-1}_{0z}\,e^{-\left(\varepsilon_{0z}/\omega_c\right)}\,,\nonumber\\
&&\\
\left.\gamma_0(\varepsilon_{0z}< 0)\right|_{T=0}
&=&-\frac{\pi\Delta^2}{2{\bf\Gamma}(2\alpha)}\left(\frac{1}{\omega_c}\right)^{2\alpha}
\left(-\varepsilon_{0z}\right)^{2\alpha-1}\,e^{\left(\varepsilon_{0z}/\omega_c\right)}.\nonumber\\
&&
\eea
\end{subequations}

Next, we consider the effect of the bath on the energy splitting.
By using the expression of $\xi(\varepsilon_{0z})$ from (\ref{eq:C_3}),
we obtain 
\bea
\left.\tilde{\varepsilon}_{0z}\right|_{T=0}&=&\left.{\varepsilon}_{0z}\times\left\lbrack
1+\frac{\xi({\varepsilon}_{0z})}{{\varepsilon}_{0z}}\right\rbrack\right|_{T=0}\nonumber\\
&=&
\varepsilon_{0z}\times
\left\lbrack
1
-\frac{\Delta^2}{\omega_c^2}\,
\frac{\sinh\left(\varepsilon_{0z}/\omega_c\right)}{\left(\varepsilon_{0z}/\omega_c\right)}\,
\frac{{\bf\Gamma}(2\alpha-2)}{{\bf\Gamma}(2\alpha)}
\right\rbrack\,.\nonumber\\
\eea
The last equation which is valid 
when $\frac{\varepsilon_{0z}}{\omega_c} \ll 1$
and $\alpha>1$ shows that $\xi({\varepsilon}_{0z})/{\varepsilon}_{0z}$ is negative
and constitute one of the principal result of this work. 
The effect of $\xi({\varepsilon}_{0z})$ is the analogue of the Lamb shift, i.e. 
the renormalization of the level splitting in atoms due to the coupling
to the electromagnetic radiations.

In thermal equilibrium, the system density matrix can be
represented in the localised eigenstates $|R\rangle$ and $|L\rangle$
of the position operator $\sigma_z =\left(|R\rangle\langle R |-|L\rangle\langle L|\right)$ as
\be
\rho = \frac{e^{\beta\hbar\varepsilon_{0z}\sigma_z/2}}
{2\cosh\left(\beta\hbar\varepsilon_{0z}/2\right)}\,.
\ee
The equilibrium values of the Bloch-vector
can be calculated from the density matrix, 
${\bf p}_{\rm st}= {\rm Tr}({\mbox{\boldmath $\sigma$}}\rho)$,
yielding
\be
\label{eq:p_equi}
{\bf p}_{\rm st}=\left({p_x}_{\rm st},{p_y}_{\rm st},{p_z}_{\rm st}\right)^T=
\left(\begin{array}{c}
0\\
0\\
\tanh(\hbar\beta\varepsilon_{0z}/2)\\
\end{array}\right)\,.
\ee
Equation (\ref{eq:p_equi}) corresponds to the 
stationary solution of the master equation 
(\ref{eq:state_equation}). 
From the rate equation (\ref{eq:pz}), it follows that the decay rate $\gamma({\varepsilon}_{0z})$
and the inhomogeneous term $\gamma_0({\varepsilon}_{0z})$
satisfies the detailed balance condition
\be
\label{eq:DB}
\frac{\gamma_0(\varepsilon_{0z})}{\gamma(\varepsilon_{0z})}
=\tanh(\hbar\beta\varepsilon_{0z}/2)\,,
\ee
which states that the process of absorption of phonons and its
inverse, the process of emission of phonons, 
occur with equal probability in thermal equilibrium and arises 
from the following quantum property 
of the thermal equilibrium correlation function 
$\langle e^{i\Omega(-t)}\,e^{i\Omega(0)}\rangle_B=
\langle e^{i\Omega(t-i\hbar\beta)}\,e^{i\Omega(0)}\rangle_B
$.

\subsection{Limits of validity of the polaron transformation\label{III-E}}

In the undriven case, the model with Hamiltonian (\ref{eq:spinbosonH})
cannot be solved analytically, in general, and there are no reliable
approximate methods which apply for a fixed (maybe weak) coupling
to the bath and for all temperatures including the very low ones.
For symmetric tunnelling $(\varepsilon_{0z}=0)$, application of perturbative 
theory in the Ohmic bath coupling leads to a non-analytical temperature dependence for
the renormalized tunnelling $\Delta_r\propto T^{\alpha}$.
At higher temperature there is a crossover from damped oscillations
to over-damped motion,~\cite{WEISS,Legget,Grabert} 
with a relaxation rate that, in the weak coupling
regime ($\alpha\ll 1$), decreases with increasing temperature,
$\gamma\propto T^{2\alpha-1}$. The singular behaviour of the 
weak coupling series shows that perturbative theories break down
at low temperature. On the other hand, the method of polaron transformation
with the resulting Hamiltonian (\ref{eq:spinbosonH_Trans}) 
is basically a perturbation theory in the tunnelling parameter $\Delta$ and 
is suitable for the strong coupling regime as we have shown in the last section.
In fact, the combination
of the polaron transformation with the second Born approximation
is equivalent to a double path integral non-interacting blip approximation
(NIBA)~\cite{Pottier}.

Recently, Vorrath {\it et al.}~\cite{Vorrath} used the combination of the polaron 
transformation with the Markov-Born approximation to derive a master equation
for the generalisation of the undriven spin-boson model to
spins greater than one-half in the strong coupling regime. 
They showed that this method is good enough if the parameters of the model, namely 
the temperature and the coupling, are limited to the case where the NIBA is valid. 
In the case of the driven spin boson model, 
the limits of the NIBA are discussed in Ref.~\cite{Grifoni}.

\section{QUANTUM OPTIMAL CONTROL PROBLEM\label{IV}}

Let the time $t$ be in the interval $[0,t_F]$ for fixed $t_F$.
The evolution of the state variable ${\bf p}(t)$ 
governed by the master equation (\ref{eq:state_equation})
depends not only on the initial state ${\bf p}(0)={\bf p}_I$
but also on the time-dependent control variable
${\mbox{\boldmath $\varepsilon$}}(t)$.
The idea now is to seek the optimal form of the 
the control field that allows for steering the Bloch vector from 
the given initial state ${\bf p}(0)={\bf p}_I$
to a desired final state at a specified time $t_F$.
Typically, it is possible to define a cost functional incorporating the
objective. Then, the goal of optimal control algorithms is to
calculate the control field which can induce a specific dynamics by 
minimizing this cost functional constraint by the state equation
~\cite{ARTHURE,Betts}, {\it i.e}, the master equation (\ref{eq:state_equation})
subject to the initial condition ${\bf p}(0)={\bf p}_I$.

\subsection{Cost functional\label{IV-A}}

For the problems of interest here, the cost functional may be written as 

\be
\label{eq:cost_Fuctional}
J = \Phi\left\lbrack{\bf p}(t_F)\right\rbrack
+ \int_0^{t_F} {\mathcal L}\left({\bf p}(t),{\mbox{\boldmath $\varepsilon$}}(t)\right)\, dt\,.
\ee
The functionals $\Phi\left\lbrack{\bf p}(t_F)\right\rbrack$ and 
${\mathcal L}\left({\bf p}(t),{\mbox{\boldmath $\varepsilon$}}(t)\right)$
account for the specific objective at the fixed target time $t_F$
and at intermediate times $t\in [0,t_F]$, respectively.
J in Eq. (\ref{eq:cost_Fuctional})
is the sum of the so-called final time cost functional 
and running cost functional.

\subsection{Pontryagin's minimum principle\label{IV-B}}
Consider the quantum optimal control problem of minimizing the cost functional
(\ref{eq:cost_Fuctional}) subject to the dynamical constraint
(\ref{eq:state_equation}):
\be
\label{eq:opt_cont_pb}
\begin{cases} 
{\mbox {min}}\left\{J = \Phi\left\lbrack{\bf p}(t_F)\right\rbrack
+ \int_0^{t_F} {\mathcal L}\left({\bf p}(t),{\mbox{\boldmath $\varepsilon$}}(t)\right)\, dt\right\}\\
\dot{\bf p}(t) = \left({\mbox{\boldmath $\varepsilon$}}_0 +{\mbox{\boldmath$\varepsilon$}}(t)\right)\times{\bf p}(t)+
{\mbox{\boldmath $\xi$}}(t)\times\left({\bf p}(t)\times{\bf n}\right)\\
\qquad\qquad-{\mbox{\boldmath $\gamma$}}(t){\bf p}(t)+{\mbox{\boldmath $\gamma$}}_0(t)\\
{\bf p}(0)={\bf p}_I\,,\quad t \in [0,t_F]\\
\end{cases}
\ee
An optimal solution of the problem (\ref{eq:opt_cont_pb}) is characterized by first order
optimality conditions in the form of the Pontryagin's minimum
principle~\cite{Pontryagin,ARTHURE,Betts,Jirari}. These conditions are formulated with 
the help of the Hamilton function that has the following form in our problem:
\bea
\label{eq:Hamilton_function}
& &{\mathcal{H}}\left(
{\bf p}(t),{\mbox{\boldmath $\lambda$}}(t),{\mbox{\boldmath $\varepsilon$}}(t)\right)=
{\mathcal L}\left({\bf p}(t),{\mbox{\boldmath $\varepsilon$}}(t)\right)+\nonumber\\
& &
{\mbox{\boldmath $\lambda$}}(t)\centerdot
\left\{
\left({\mbox{\boldmath $\varepsilon$}}_0 +{\mbox{\boldmath$\varepsilon$}}(t)\right)\times{\bf p}(t)+
{\mbox{\boldmath $\xi$}}(t)\times\left({\bf p}(t)\times{\bf n}\right)
\right.\nonumber\\
& &
\left.~~~~
-{\mbox{\boldmath $\gamma$}}(t){\bf p}(t)+{\mbox{\boldmath $\gamma$}}_0(t)
\right\}\,,
\eea
where ${\bf \lambda(t)}$ is called the adjoint or the co--state
variable, which is a Lagrange multiplier introduced to implement
the constraint and thereby to render the variables ${\bf p}(t)$ and ${\mbox{\boldmath$\varepsilon$}}(t)$ 
independent. Following a variation in analogy to Hamilton's variation principle of classical mechanics, the 
necessary first order optimality conditions result in a two-point boundary value problem:
\be
\label{eq:TWBVP}
\begin{cases}
\dot{\bf p}(t)= \frac{\partial{\mathcal{H}}\left({\bf p}(t),{\mbox{\boldmath $\lambda$}}(t),{\mbox{\boldmath $\varepsilon$}}(t)\right)}
{\partial{\mbox{\boldmath $\lambda$}}(t)},\quad t\in[0,t_F]\\
\\
\dot{\mbox{\boldmath $\lambda$}}(t) = 
-\frac{\partial{\mathcal{H}}\left({\bf p}(t),{\mbox{\boldmath $\lambda$}}(t),{\mbox{\boldmath $\varepsilon$}}(t)\right)}
{\partial{\bf p}(t)},\qquad t\in[0,t_F]\\
\\
{\bf p}(0)={\bf p}_I,\qquad {\mbox{\boldmath $\lambda$}}(t_F)=\frac{\partial\Phi\left\lbrack{\bf p}(t_F)\right\rbrack}{\partial{\bf p}(t_F)}\\
\\
0=\frac{\partial{\mathcal{H}}\left({\bf p}(t),{\mbox{\boldmath $\lambda$}}(t),{\mbox{\boldmath $\varepsilon$}}(t)\right)}
{\partial{\mbox{\boldmath $\varepsilon$}}(t)},\quad t\in[0,t_F]
\end{cases}
\ee
where the last condition is equivalent to the vanishing of the 
first variation of the cost functional,{\it i.e}, $\delta J =0$.

The minimum principle requires the solution of complicated nonlinear algebraic equations,                                                 
namely,  the optimality condition 
${\partial{\mathcal{H}}}/{\partial{\mbox{\boldmath $\varepsilon$}}}=0$,
which can only be solved in an iterative manner.
The present optimal control problem (\ref{eq:opt_cont_pb}) is not singular because 
${\partial^2{\mathcal{H}}}/{\partial{\mbox{\boldmath $\varepsilon$}}^2}\not= 0$, since
${\mbox{\boldmath $\xi$}}(t)$, ${\mbox{\boldmath $\gamma$}}(t)$ and ${\mbox{\boldmath $\gamma$}}_0(t)$
depend nonlinearly on ${\mbox{\boldmath $\varepsilon$}}(t)$ regardless of the choice 
of the running cost ${\mathcal L}$.
Applying then the implicit function theorem one concludes that the optimality condition 
may have one solution, {\it i.e}, 
${\mbox{\boldmath $\varepsilon$}}(t)=\left({\bf p}(t), {\mbox{\boldmath $\lambda$}}(t)\right)$
or more solutions which are only ''locally unique''.
Here we apply the gradient method as an iterative scheme for
solving (\ref{eq:opt_cont_pb}). 

Let us now explicitly compute the gradient of the cost functional
with respect to the control field. For this we first write the equation of motion for the adjoint state
${\mbox{\boldmath $\lambda$}}$. Eqs.~(\ref{eq:Hamilton_function}) and (\ref{eq:TWBVP} lead to
\bea
\label{eq:costate_equation}
&&\dot{\mbox{\boldmath $\lambda$}}(t) = 
-\frac{\partial{\mathcal L}\left({\bf p}(t),{\mbox{\boldmath $\varepsilon$}}(t)\right)}{\partial{\bf p}(t)}
+
\left({\mbox{\boldmath $\varepsilon$}}_0 +
{\mbox{\boldmath$\varepsilon$}}(t)\right)\times{\mbox{\boldmath $\lambda$}}(t)\nonumber\\
&&
\qquad\qquad
-
\left({\mbox{\boldmath $\xi$}}(t)\times{\mbox{\boldmath $\lambda$}}(t)\right)\times{\bf n}
+{\mbox{\boldmath $\gamma$}}^T(t){\mbox{\boldmath $\lambda$}}(t).
\eea
or in components form
\begin{subequations}
\bea
\dot \lambda_x(t) &=& -\frac{\partial{\mathcal L}}{\partial{p_x}(t)}
+\left(\varepsilon_{0z} + \varepsilon_z(t)+\xi(t)\right)\lambda_y(t)\,,\\
&&\nonumber\\
\dot\lambda_y(t)&=& -\frac{\partial{\mathcal L}}{\partial{p_y}(t)}
-\left(\varepsilon_{0z} + \varepsilon_z(t)\right)\lambda_x(t)+\gamma(t)\lambda_y(t)\,,\nonumber\\
&&\\
\dot\lambda_z(t)&=&-\frac{\partial{\mathcal L}}{\partial{p_z}(t)}+\gamma(t)\lambda_z(t)\,.
\eea
\end{subequations}

Within the Pontryagin's Minimum Principle, the variation 
of the cost functional (\ref{eq:cost_Fuctional}) reads
%
%
\bea
\label{eq:delta_cost_Fuctional}
\delta J &=& \int_0^{t_F}\frac{\delta J}
{\delta{\mbox{\boldmath $\varepsilon$}}}\centerdot\delta{\mbox{\boldmath $\varepsilon$}}dt
\label{eq:dJdEpsilon}
=\int_0^{t_F}
\frac{\partial{\mathcal{H}}\left({\bf p}(t),{\mbox{\boldmath $\lambda$}}(t),{\mbox{\boldmath $\varepsilon$}}(t)\right)}
{\partial{\mbox{\boldmath $\varepsilon$}}(t)}\centerdot\delta{\mbox{\boldmath $\varepsilon$}}(t)dt\nonumber\\
&=&
\int_0^{t_F}
\left\lbrack
\frac{\partial{\mathcal L}\left({\bf p}(t),{\mbox{\boldmath $\varepsilon$}}(t)\right)}{\partial{\mbox{\boldmath$\varepsilon$}}(t)}+
\left({\mbox{\boldmath $\lambda$}}(t)\times{\bf p}(t)\right)
\right\rbrack\centerdot\delta{\mbox{\boldmath $\varepsilon$}}(t)dt\nonumber\\
&&
+\int_0^{t_F}
\frac{\partial}{\partial{\mbox{\boldmath$\varepsilon$}}(t)}
\left\lbrack
{\mbox{\boldmath $\lambda$}}(t)\centerdot
\{
{\mbox{\boldmath $\xi$}}(t)\times\left({\bf p}(t)\times{\bf n}\right)\}
\right\rbrack\centerdot\delta{\mbox{\boldmath $\varepsilon$}}(t)dt\nonumber\\
&&
+\int_0^{t_F}
\frac{\partial}{\partial{\mbox{\boldmath$\varepsilon$}}(t)}
\left\lbrack
{\mbox{\boldmath $\lambda$}}(t)\centerdot
\left\{
-{\mbox{\boldmath $\gamma$}}(t){\bf p}(t)+{\mbox{\boldmath $\gamma$}}_0(t)
\right\}
\right\rbrack\centerdot\delta{\mbox{\boldmath $\varepsilon$}}(t)dt\,.\nonumber\\
&&
\eea
%
%
%
%
The last equation enables us to compute the gradient of the cost functional 
with respect to the control field.

We may summarize the whole procedure for computing the gradient of the cost functional as follows: 
i) for a given control field
${\mbox{\boldmath $\varepsilon$}}(t)$, we first solve the state equation (\ref{eq:state_equation})
forward in time,
ii) the solution obtained for ${\bf p}$ is then used for the backward integration 
of the adjoint equation (\ref{eq:costate_equation}),
iii) with ${\bf p}$ and ${\mbox{\boldmath $\lambda$}}$ obtained we compute the gradient.
\begin{table}[t]
\caption{Best-fit parameters for a model defined by Eq.~(\ref{eq:fit}).
$\Delta$ is the unit of the amplitudes $A_i$,
$\Delta/2\pi$ of the frequencies $\nu_i$ and $\mbox{rd}$  of the phases
$\phi_i$. $\chi^2=36.43$.}
\begin{ruledtabular}
\begin{tabular}{ccc}
Parameters & Fit & Error\\
\hline
$\nu$     &   0.354089322 & 2.431734350$\times 10^{-5}$\\
\hline
$A_1$  	  &   2.38068007  & 6.047635201$\times 10^{-3}$\\
$\phi_1$  & -1.83291567   &       2.978756342$\times 10^{-3}$\\
\hline
$A_2$     &2.79811789      &   6.033930931$\times 10^{-3}$\\
$\phi_2$  & 0.960902574    &      3.718809069$\times 10^{-3}$\\
\hline
$A_3$     &-0.909244535      &    6.041502784$\times 10^{-3}$\\
$\phi_3$  & 0.65358378    &     8.091472983$\times 10^{-3}$\\
\hline
$A_4$     & 0.448104715      &   6.048448725$\times 10^{-3}$\\
$\phi_4$  & -5.91914473    &      1.480647739$\times 10^{-2}$\\
\hline
$A_5$     & 0.144629845      &    6.054115543$\times 10^{-3}$\\
$\phi_5$  & -9.41113677    &      4.260009231$\times 10^{-2}$\\
\hline
$A_6$     & -0.0758863206      &   6.060940693$\times 10^{-3}$\\
$\phi_6$ &  2.94649922    &      8.029036597$\times 10^{-2}$\\
\hline
$A_7$     & -0.0170884169      &    6.059845970$\times 10^{-3}$\\
$\phi_7$  & -0.882323605    &     3.542289951$\times 10^{-1}$
\end{tabular}
\end{ruledtabular}
\label{Tab1}
\end{table}
\subsection{Discretization\label{sect_Dis}}

By an appropriate discretization scheme, the above infinite dimensional 
constraint optimal control problem can be transformed into a finite 
dimensional optimization approximation~\cite{ARTHURE,Betts}. 
For this purpose, we subdivide the time interval $[t_1=0,t_F]$ 
using
\be
t_{j+1}=t_j+\Delta t,\quad j=1,\ldots,M-1\quad{\mbox{with}}\quad\Delta t = t_F/M
\ee
%
%
%

The values of the state, the adjoint and the control 
are evaluated at the mesh points $t_j$ 
\be
\left({\bf p},{\mbox{\boldmath $\lambda$}},{\mbox{\boldmath$\varepsilon$}}\right)=
\left({\bf p}_1,\ldots, {\bf p}_M,
{\mbox{\boldmath $\lambda$}}_1,\ldots, {\mbox{\boldmath $\lambda$}}_M,
{\mbox{\boldmath$\varepsilon$}}_1,\ldots, {\mbox{\boldmath$\varepsilon$}}_M\right)^T
\, \in {\mathbb{R}}^{9M}
\ee
where the following notation 
${\bf p}(t_j):={\bf p}_j$, 
${\mbox{\boldmath$\lambda$}}(t_j):={\mbox{\boldmath $\lambda$}}_j$
and 
${\mbox{\boldmath$\varepsilon$}}(t_j):={\mbox{\boldmath $\varepsilon$}}_j$
is used.

Adopting the Euler scheme for 
solving the state equation (\ref{eq:state_equation}) and the adjoint equation
(\ref{eq:costate_equation})
and by applying the Riemann-rule integration
to the cost functional (\ref{eq:cost_Fuctional})
and to its variation (\ref{eq:delta_cost_Fuctional}), 
we obtain the main tool for solving the time-discrete formulation of the quantum optimal control problem 
defined in Eq.~(\ref{eq:opt_cont_pb})
%
%
\begin{itemize}
\item{state equation}
\bea
\label{eq:state_equation_dis}
&&{\bf p}_{j+1} = {\bf p}_{j}+\Delta t\left\lbrack
\left(
{\mbox{\boldmath $\varepsilon$}}_0 +{\mbox{\boldmath$\varepsilon$}}_{j}\right)\times{\bf p}_j+
{\mbox{\boldmath $\xi$}}_j
\times
\left(
{\bf p}_j\times{\bf n}
\right)\right\rbrack\nonumber\\
&&\qquad\qquad
-\Delta t
\left\lbrack{\mbox{\boldmath $\gamma$}}_j{\bf p}_j+{{\mbox{\boldmath $\gamma$}}_0}_j\right\rbrack
\eea
for $j=1,\ldots,M-1$ with ${\bf p}_1={\bf p}_{I}$
\item{adjoint equation}
\bea
\label{eq:costate_equation_dis}
&&{\mbox{\boldmath $\lambda$}}_{j-1} ={\mbox{\boldmath $\lambda$}}_{j}
-\Delta t\left\lbrack
-\frac{\partial{\mathcal L}\left({\bf p}_j,{\mbox{\boldmath $\varepsilon$}}_j\right)}{\partial{\bf p}_j}
+
\left({\mbox{\boldmath $\varepsilon$}}_0 +
{\mbox{\boldmath$\varepsilon$}}_j\right)\times{\mbox{\boldmath $\lambda$}}_j
\right\rbrack
\nonumber\\
&&
\qquad\qquad
-\Delta t\left\lbrack
\left({\mbox{\boldmath $\xi$}}_j\times{\mbox{\boldmath $\lambda$}}_j\right)\times{\bf n}
+{\mbox{\boldmath $\gamma$}}^T_j{\mbox{\boldmath $\lambda$}}_j
\right\rbrack
\eea
for $j=M,\ldots,2$ with ${\mbox{\boldmath $\lambda$}}_M=\frac{\partial\Phi\left({\bf p}_M\right)}{\partial{\bf p}_M}$.
\item{cost functional}
\be
\label{eq:cost_Fuctional_dis}
J = \Phi\left({\bf p}_M\right)
+\Delta t\sum_{j=1}^M\,
{\mathcal L}\left({\bf p}_j,{\mbox{\boldmath $\varepsilon$}}_j\right)
\ee
\end{itemize}
\begin{itemize}
\item{variation of the cost functional}
\bea
&&\Delta J = \Delta t \sum_{j=1}^{M}\,
\left\lbrack
\frac{\partial{\mathcal L}\left({\bf p}_j,{\mbox{\boldmath $\varepsilon$}}_j\right)}
{\partial{\mbox{\boldmath$\varepsilon$}}_j}+
\left({\mbox{\boldmath $\lambda$}}_j\times{\bf p}_j\right)
\right\rbrack
\centerdot\Delta{\mbox{\boldmath $\varepsilon$}}_j\nonumber\\
&&\qquad
+\Delta t \sum_{j,k=1}^{M}\,
\frac{\partial}{\partial{\mbox{\boldmath $\varepsilon$}}_k}
\left\lbrack
{\mbox{\boldmath $\lambda$}}_j\centerdot
\left\{
{\mbox{\boldmath $\xi$}}_j\times\left({\bf p}_j\times{\bf n}\right)
\right\}
\right\rbrack
\centerdot\Delta{\mbox{\boldmath $\varepsilon$}}_k\nonumber\\
&&
+\Delta t \sum_{j,k=1}^{M}\,
\frac{\partial}{\partial{\mbox{\boldmath $\varepsilon$}}_k}
\left\lbrack
{\mbox{\boldmath $\lambda$}}_j\centerdot
\left\{
-{\mbox{\boldmath $\gamma$}}_j{\bf p}_j+
{{\mbox{\boldmath $\gamma$}}_0}_j
\right\}
\right\rbrack
\centerdot\Delta{\mbox{\boldmath $\varepsilon$}}_k\nonumber\\
&&
\eea
\item{gradient of the cost functional}
\bea
\label{eq:3d_gradient}
&&\frac{\partial J}{\partial{\mbox{\boldmath$\varepsilon$}}_i}=
\Delta t\left\lbrack
\frac{\partial{\mathcal L}\left({\bf p}_i,{\mbox{\boldmath $\varepsilon$}}_i\right)}
{\partial{\mbox{\boldmath$\varepsilon$}}_i}+
{\mbox{\boldmath $\lambda$}}_i\times{\bf p}_i
\right\rbrack\nonumber\\
&&
+\Delta t \sum_{j=1}^{M}\,
\frac{\partial}{\partial{\mbox{\boldmath $\varepsilon$}}_i}
\left\lbrack
{\mbox{\boldmath $\lambda$}}_j\centerdot
\left\{
{\mbox{\boldmath $\xi$}}_j\times\left({\bf p}_j\times{\bf n}\right)
-{\mbox{\boldmath $\gamma$}}_j{\bf p}_j+
{{\mbox{\boldmath $\gamma$}}_0}_j
\right\}
\right\rbrack\nonumber\\
&&
\eea
for $i=1,\ldots,M$.
\end{itemize}

The projection of (\ref{eq:3d_gradient}) on the z-axis of the Bloch sphere leads to
\bea
\label{eq:dj_de_k}
&&\frac{\partial J}{\partial\varepsilon_{zi}}=\Delta t
\left\lbrack
\frac{
\partial{\mathcal L}
\left(
p_{xi},p_{yi},p_{zi},\varepsilon_{zi}
\right)
}{\partial\varepsilon_{zi}}
+\left(
\lambda_{xi}p_{yi}-\lambda_{yi}p_{xi}
\right)
\right\rbrack
\nonumber\\
&&\qquad\qquad
-\Delta t \sum_{j=1}^{M}\,
\lambda_{yj}
\left\lbrack
\frac{\partial\xi_j}{\partial\varepsilon_{zi}}p_{xj}
+\frac{\partial\gamma_j}{\partial\varepsilon_{zi}}p_{yj}
\right\rbrack
\nonumber\\
&&\qquad\qquad
+\Delta t \sum_{j=1}^{M}\,
\lambda_{zj}
\left\lbrack
-\frac{\partial\gamma_j}{\partial\varepsilon_{zi}}p_{zj}
+\frac{\partial\gamma_{0j}}{\partial\varepsilon_{zi}}
\right\rbrack
\eea
for $i=1,\ldots,M$

The matrices ${\partial\xi_{j}}/{\partial\varepsilon_i}$,
${\partial\gamma_{j}}/{\partial\varepsilon_i}$ and ${\partial\gamma_{0j}}/{\partial\varepsilon_i}$
of size $M\times M$ are not diagonal because 
in the time-continuous problem $\xi(t)$, $\gamma(t)$ and $\gamma_0(t)$
at the current time $t$ depend on the control field at $t'\le t$
via the function (\ref{eq:field_effect_Func}).
\begin{figure}
\includegraphics[width=7cm,angle=-90]{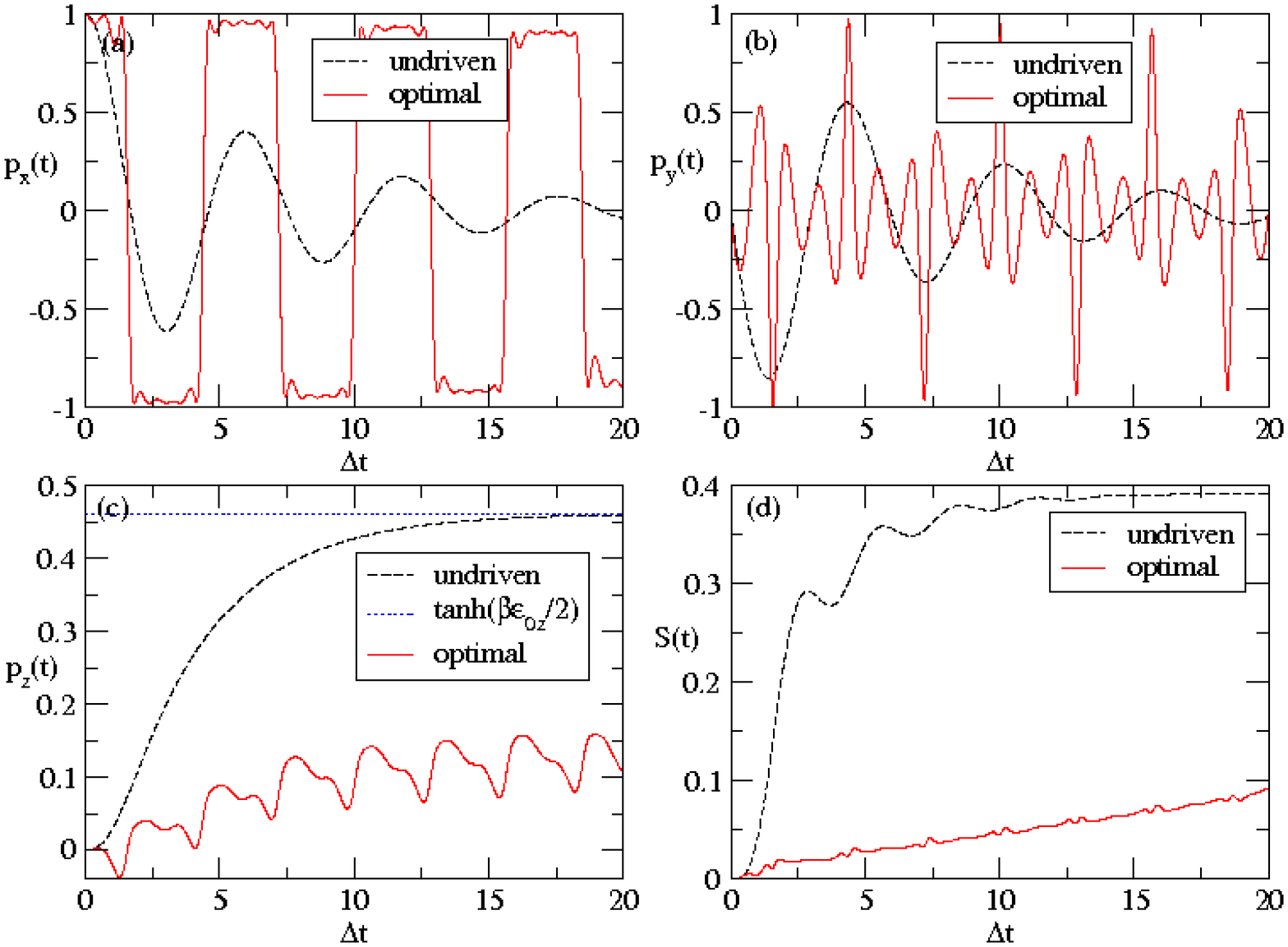}
\caption{Implementation of the Z-gate with 
${\bf p}_I=(1,0,0)^T$ and ${\bf p}_d=(-1,0,0)^T$.
Depicted are the Bloch vector
${\bf p}=\left(p_x, p_y, p_z\right)^T$ and the linear
entropy $S = \frac{1}{2}\left(1-||{\bf p}||^2\right)$ as a function of time 
for undriven case and for 
driven by the optimal control field obtained by the conjugate gradient method.
(a), (b) and (c) show, respectively, the results for $p_x$ , $p_y$ and $p_z$ 
while (d) show the results for $S$.
The parameters used are $\alpha=0.2$, $\varepsilon_{0z}=\Delta$,
$\omega_c=20\Delta$ and $k_BT=\beta^{-1} = \hbar\Delta$.
The final time is set as $t_F= 20/\Delta$
and the chosen time step is $10^{-2}/\Delta$ corresponding 
to $M=2\times 10^3$ as the number of mesh points,
{\it i.e}, the dimension of the optimal control problem.}
\label{fig:fig1}
\end{figure}
\begin{figure}
\includegraphics[width=7cm,angle=-90]{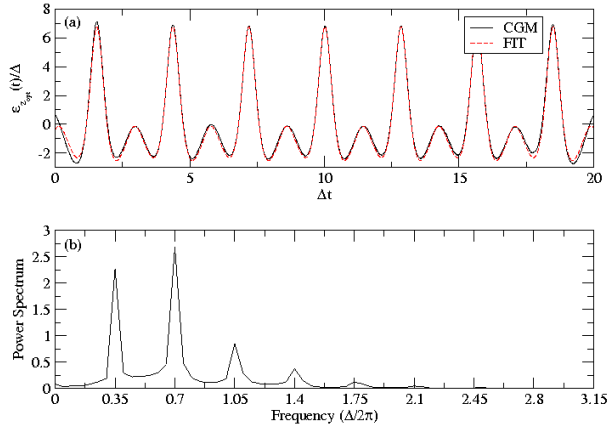}
\caption{Implementation of the Z-gate with 
${\bf p}_I=(1,0,0)^T$ and ${\bf p}_d=(-1,0,0)^T$.
The upper panel (a) shows the optimal control field selected by the conjugate
gradient method (CGM) vs. time while the lower panel (b) shows its power spectrum.
A comparison with the model fit defined by Eq.~(\ref{eq:fit}) is also shown 
in the upper panel (a). Parameters as in Fig.~\ref{fig:fig1}.}
\label{fig:fig2}
\end{figure}

\subsection{Numerical method}

The set of equations needed to solve the optimal control problem (\ref{eq:opt_cont_pb}) are
the discrete-time versions of the cost functional $J(\varepsilon_{z1}\ldots,\varepsilon_{zM})$ 
defined in Eq.~(\ref{eq:cost_Fuctional_dis}),
the equation of motion for the state and the adjoint variables 
given by Eqs.~(\ref{eq:state_equation_dis}) and (\ref{eq:costate_equation_dis}),
respectively, and the gradient of the cost functional 
${\nabla}J(\varepsilon_{z1}\ldots,\varepsilon_{zM})$
in the form of Eq.~(\ref{eq:dj_de_k}).

If we can compute the cost functional and its
gradient at arbitrary points 
$\varepsilon_{z}=(\varepsilon_{z1}\ldots,\varepsilon_{zM})^T\in{\mathbb{R}}^M$, 
the general form of the gradient algorithm 
for minimization is as follows~\cite{Frederic}
\begin{enumerate}
\item Initialization:
the initial guess ${\varepsilon_z^1}\in{\mathbb{R}}^M$ and 
the stopping tolerance tol $> 0$ are given; set i=1.
\item Stopping test: 
\begin{itemize}
\item integrate the state equation forward in time to find ${\bf p}$.
\item integrate the adjoint equation backward in time to find ${\mbox{\boldmath $\lambda$}}$
\item compute the gradient $\nabla J(\varepsilon_z^1)$; 

\nobreak{if $\left|\nabla J(\varepsilon_z^1)\right| \leq $ tol stop.}
\end{itemize}
\item Computing the direction: compute the descent direction
${d^i}\in{\mathbb{R}}^M$ defined by $ \nabla J(\varepsilon_z^i)\centerdot{d^i} < 0$.
\item Line-search: find an appropriate stepsize $\mu^i > 0$
satisfying $J(\varepsilon_z^i+ \mu^i\,d^i) < J(\varepsilon_z^i)$
\item Loop: \nobreak{
set $\varepsilon_z^{i+1}=\varepsilon_z^i+ \mu^i\,{d^i}$;
increase i by 1 and go to 2.
}
\end{enumerate}

In the work described here, the optimization of 
the cost functional is performed by using 
the subroutine FRPRMN of the Numerical Recipes package~\cite{NRC}
which implements the conjugate gradient method as a 
variant of the above descent algorithm .
We also used the subroutine DMNG of PORT library~\cite{port}
implementing the quasi-Newton method.
These two iterative methods of optimization are very popular.
Both of them require the gradient but differ
in the calculation of the descent direction.

The equations of motion for the state and the adjoint variables
are forward and backward initial value problems, respectively.
We solved them using the Euler scheme or a Runge-Kutta scheme 
which requires the values for the control field only at a grid point
(see Sec.~\ref{sect_Dis}).  
Evaluation of the state and the adjoint variables involves an extensive computation
of the time-dependent rates which are given by an integral over time of a 
rapidly oscillating functions (see Eqs.~(\ref{eq:gamma_yx}), (\ref{eq:gamma_zz}), (\ref{eq:gamma_z0}),
(\ref{eq:Q_1}) and (\ref{eq:Bis_Q_2})).
The numerical evaluation of the rate functions and their derivatives with
respect to the control field involved in the computation of the gradient
are performed using a Gauss quadrature suitable
for an integration of rapidly oscillating functions.
\begin{figure}
\includegraphics[width=6cm,angle=-90]{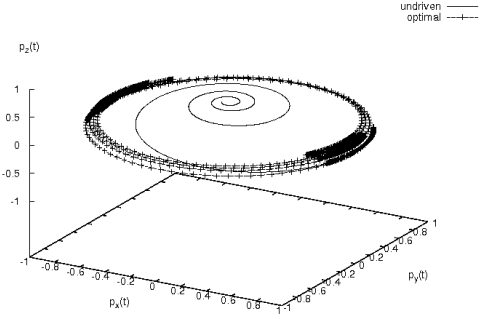}
\caption{Implementation of the Z-gate with 
${\bf p}_I=(1,0,0)^T$ and ${\bf p}_d=(-1,0,0)^T$.
The 3-dimensional plot of the the Bloch vector
${\bf p}=\left(p_x, p_y, p_z\right)^T$
for undriven case and for 
driven by the optimal control field
obtained by the conjugate gradient method
is presented. Parameters as in Fig.~\ref{fig:fig1}.}
\label{fig:fig3}
\end{figure}
\begin{figure}
\includegraphics[width=7cm,angle=-90]{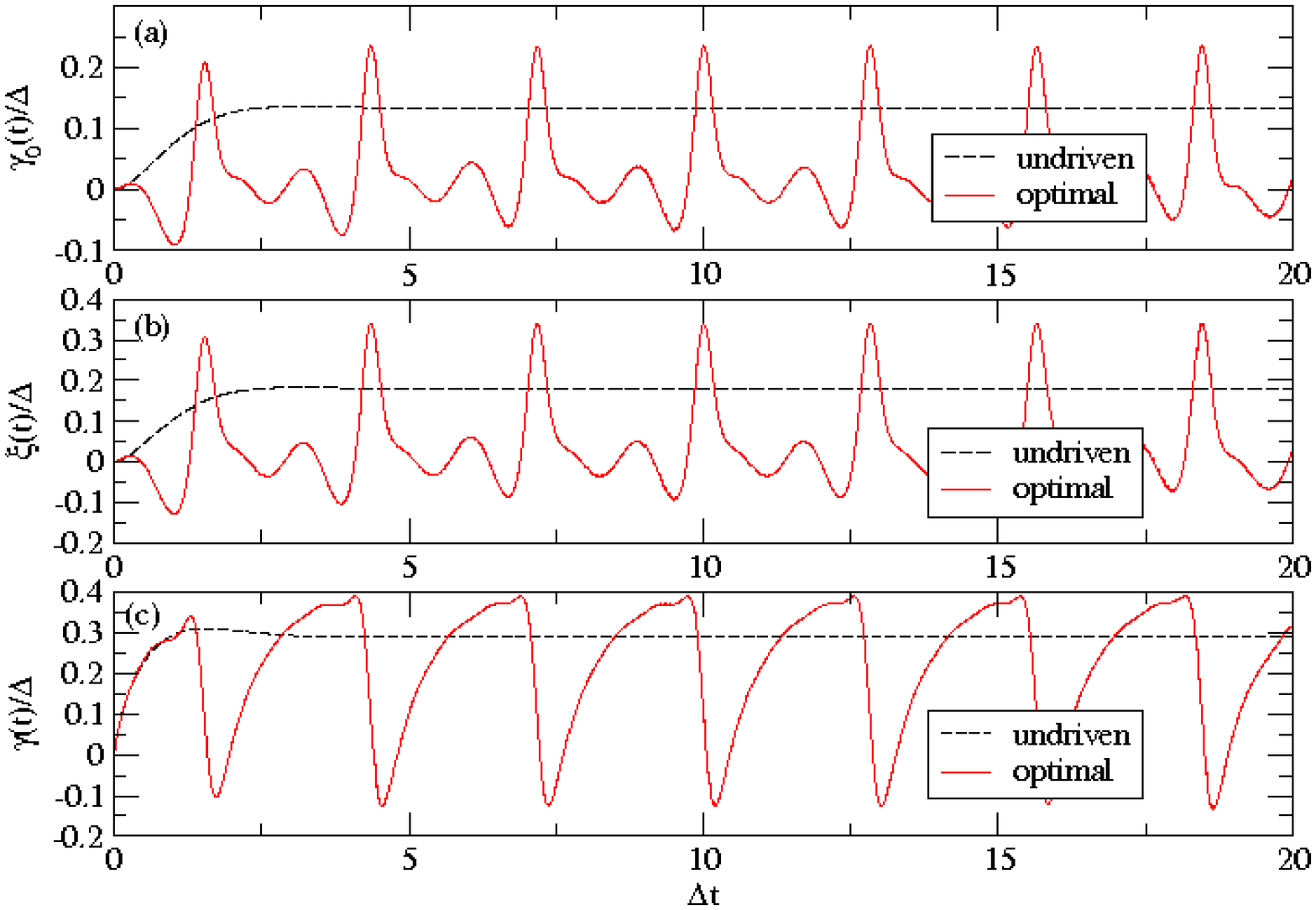}
\caption{Implementation of the Z-gate with 
${\bf p}_I=(1,0,0)^T$ and ${\bf p}_d=(-1,0,0)^T$.
Depicted are the decay rate $\gamma$,
the Lamb shift $\xi$ and the inhomogeneous term $\gamma_0$ 
as a function of time for undriven case and driven by the optimal control 
field selected by the conjugate
gradient method. (a), (b) and (c) show, respectively, the results for $\gamma_0$, $\xi$
and $\gamma$. Parameters as in Fig.~\ref{fig:fig1}.}
\label{fig:fig4}
\end{figure}

\section{NUMERICAL RESULTS\label{V}}
\subsection{Z gate}

As a first application of the quantum optimal control theory developed in Sec.~\ref{IV},
we consider the action of the Z gate
\be
Z
= \left(\begin{array}{c c}
      1 &0\\
      0 &-1 
\end{array}\right),
\ee
which leaves $|0\rangle$  unchanged, and flips $|1\rangle$ to $-|1\rangle$. Its application to 
the initial state $|\psi\rangle=\left(|0\rangle+|1\rangle\right)/\sqrt 2$
leads to $|\psi\rangle'=\left(|0\rangle-|1\rangle\right)/\sqrt 2$.
In term of the density matrix or the Bloch vector, we have for this particular state 
\be
\begin{CD}
\rho = \frac{1}{2}
\left(\begin{array}{c c}
      1 &1\\
      1& 1 
\end{array}\right) 
@>Z>>
\rho' = \frac{1}{2}
\left(\begin{array}{c c}
      1 &-1\\
      -1& 1 
\end{array}
\right)\\
@VVV    @VVV\\
{\bf p}=
\left(\begin{array}{c}
1\\
0\\
0\\
\end{array}\right) 
@>Z>> 
{\bf p}'=
\left(\begin{array}{c}
-1\\
0\\
0\\
\end{array}\right) 
\end{CD}
\ee
The action of the dissipative Z-gate is phrased as an optimization problem.
At time $t_I=0$ the two-level system (qubit) is prepared in the  
initial state ${\bf p}_I=(1,0,0)^T$.
Our objective is to bring it into the desired state  
${\bf p}_d=(-1,0,0)^T$ at time $t=t_F$. In this case, we need to minimize the deviation
of the state of the system at final time ${\bf p}(t_F)$ from the desired state
${\bf p}_d$. The cost functional chosen for this task is
\be
\label{eq:Z_J}
J =\frac{1}{2}\Vert {\bf p}(t_F) -{\bf p}_d\Vert^2~~, 
\ee
corresponding to the running cost functional  
${\mathcal L}\left({\bf p}(t),{\mbox{\boldmath $\varepsilon$}}(t)\right)=0$ for all $t\in [0,t_F]$
and to the final cost functional $\Phi\left\lbrack{\bf p}(t_F)\right\rbrack=\frac{1}{2}\Vert {\bf p}(t_F) -{\bf p}_d\Vert^2$
in Eq.~(\ref{eq:cost_Fuctional}).
The cost functional defined in Eq.~(\ref{eq:Z_J}) requires 
${\mbox{\boldmath $\lambda$}}(t_F)={\bf p}(t_F) -{\bf p}_d$
as the initial condition in Eq.~(\ref{eq:costate_equation}) 
for the backward integration of the adjoint state variables
${\mbox{\boldmath $\lambda$}}$.

Fig.~\ref{fig:fig1} shows the components of the Bloch vector versus time and 
the evolution of the linear entropy defined by~\cite{Breuer}
\be
S(t) = (1-\Vert{\bf p}(t)\Vert^2)/2.
\ee
The dashed lines give the result for the case of zero control $\varepsilon(t)=0$.   
The solid lines give the results for the optimum field which was 
obtained by starting from a zero initial field and allowing 20 iterations. 
Fig.~\ref{fig:fig2} shows the optimal field versus time, as well as its power spectrum.   
It can be seen that the selected field performs several abrupt switch operations between initial state 
${\bf p}_I=(1,0,0)^T$  and target state 
${\bf p}_d=(-1,0,0)^T$ to arrive at the target state at time $t_F$.   
In principle the Z--gate operation is completed at approximately time $t=2.5\Delta$.  
However, here we are interested in 
preventing the decrease of the Bloch vector over a prolonged period of time. 
The physical interpretation to the selected solution is the following: 
inspection of the kinetic equations for the Bloch vector Eqs.(\ref{eq:px}),
(\ref{eq:py} and (\ref{eq:px})
shows that a static field ${\varepsilon_z}_{\mbox{opt}}(t)=-\varepsilon_{0z}$  makes
$(p_x,0,0)^T$, for $-1\leq p_x\leq +1$ a stable (``decoherence--free") subspace of the driven system.   
In this optimization run, the gradient selected a multiple switching version, whereby the system 
is, approximately,  switched between the decoherence--free states ${\bf p}(t)=(1,0,0)^T$ and ${\bf p}(t)=(-1,0,0)^T$.
Dissipation essentially is initiated during the first switching operation 
when there is a small build--up in $p_y$ and $p_z$ component, 
as can be seen by inspection of Fig~\ref{fig:fig1}(d) showing the linear entropy of the system.   
The latter increases almost linearly with time, however, at a greatly reduced rate when compared to the 
time--evolution of the undriven system.
The situation is complicated because 
$p_z$ has a thermal equilibrium state at around $0.46$.
As one sees in Fig.~\ref{fig:fig1}(c), the optimal field succeeds repeatedly in driving the $p_z(t)$ back towards zero.    
The evolution of the 3-d Bloch vector is shown in Fig.~\ref{fig:fig3}.     
While the control--free evolution rapidly spirals towards the thermal equilibrium state 
${\bf p}_{\mbox{st}}=(0,0,\tanh(\hbar\beta\varepsilon_{0z}/2))^T$
the selected optimum control field is able to stabilize the Bloch vector
and eventually drive it very near to the target state ${\bf p}_d=(-1,0,0)^T$. 
\begin{figure}
\includegraphics[width=7cm,angle=-90]{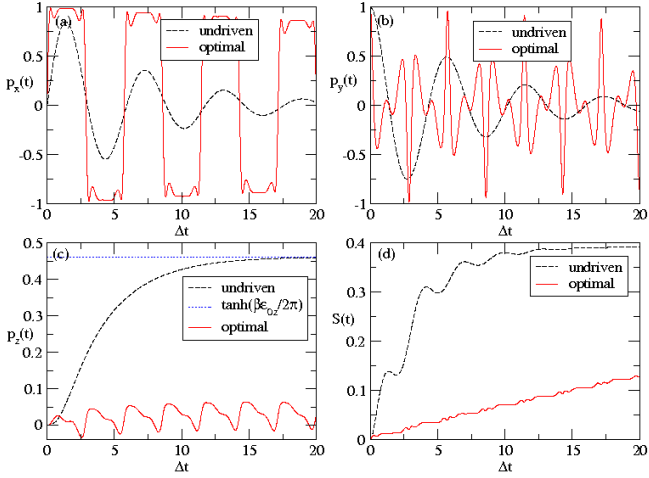}
\caption{Implementation of the Z-gate with 
${\bf p}_I=(0,1,0)^T$ and ${\bf p}_d=(0,-1,0)^T$.
Depicted are the Bloch vector
${\bf p}=\left(p_x, p_y, p_z\right)^T$ and the linear
entropy $S = \frac{1}{2}\left(1-||{\bf p}||^2\right)$ as a function of time 
for undriven case and for 
driven by the optimal control field obtained by the conjugate gradient method.
(a), (b) and (c) show, respectively, the results for $p_x$ , $p_y$ and $p_z$ 
while (d) show the results for $S$.
Parameters as in Fig.~\ref{fig:fig1}}
\label{fig:bfig1}
\end{figure}
\begin{figure}
\includegraphics[width=7cm,angle=-90]{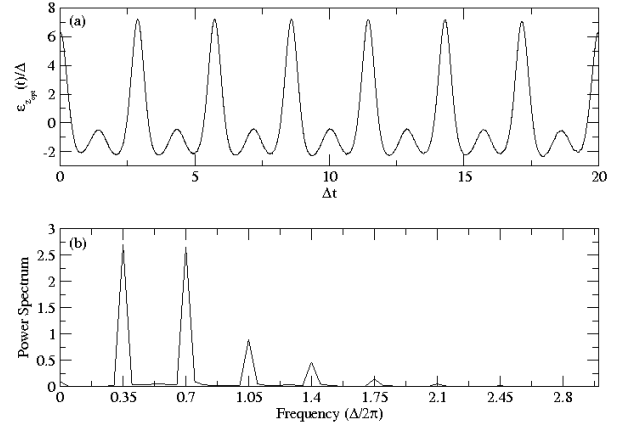}
\caption{Implementation of the Z-gate with 
${\bf p}_I=(0,1,0)^T$ and ${\bf p}_d=(0,-1,0)^T$.
The upper panel (a) shows the optimal control field selected by the conjugate
gradient method (CGM) vs. time while the lower panel (b) shows its power spectrum.}
\label{fig:bfig2}
\end{figure}

Fig~\ref{fig:fig2}(a) displays the time evolution of the selected optimal field.   
The repeated switching of the Bloch vector is achieved by a nearly periodic field.  
The essence of the Z--gate operation is more or less contained in one period.   
The electric field oscillates about the value $\varepsilon_{\mbox opt}(0)\sim-\varepsilon_{0z}$ to 
trap the system in state ${\bf p}=(\pm 1,0,0)$.  The switching is performed by a positive pulse 
which is optimized to rotate the Bloch vector into state ${\bf p}=(\mp 1,0,0)^T$. 
Then the field goes negative again to trap the system in this state.   Performing 
more iterations will smoothen the oscillation about $\varepsilon_{\mbox opt}(0)\sim-\varepsilon_{0z}$ and reduce 
the slope in the rise of the linear entropy. 
The analysis of Figs.~\ref{fig:fig1} and \ref{fig:fig2} shows that
the small oscillations of the control field about the value $-\varepsilon_{0z}$ 
between two switching operations (two positive pulses)
are reflected in the time evolution of the Bloch vector.

The influence of the control--field on the dissipative part of the kinetic equations
and the energy renormalization  is displayed 
in Fig.~\ref{fig:fig4} showing $\gamma_0$, $\gamma$, and $\xi$ versus time for the driven and undriven case.
The periodic structure of the optimal control field manifests itself in both of them. 
The renormalization term $\xi$ and $\gamma_0$ resemble, essentially, 
a shifted and rescaled version of the control field itself.   
In this fashion they optimize support for the action of the electric field, in particular, 
when the latter rises to perform a switching operation.  
The minima of the relaxation rate $\gamma$, on the other hand,  
occur when the control field becomes large. 
In this way, dissipation during the switching process is minimized.

Fig~\ref{fig:fig2}(b) displays the power spectrum of the selected optimal control field
showing seven pronounced peaks at near equidistant frequencies.
So, the selected optimal control field can be approximated by
\be
\label{eq:fit}
{\varepsilon_z}^{{\small\mbox{FIT}}}(t)=\sum_{n=1}^7\, A_n\sin(2\pi n\nu t + \phi_n)
\ee
depending on $15$ adjustable parameters which we determine
using a nonlinear least square method consisting of minimizing
the $\chi^2$ merit function defined by
\be
\chi^2 = \frac{1}{2}\sum_{i=1}^{M}
\left\lbrack
{\varepsilon_z}^{{\small\mbox{CGM}}}(t_i) - {\varepsilon_z}^{{\small\mbox{FIT}}}(t_i)\right\rbrack^2
\ee
where $M =2000$ is the number of mesh points and ${\varepsilon_z}^{{\small\mbox{CGM}}}(t_i)$
is the optimal control field  shown in Fig~\ref{fig:fig2}(a), solid line.
The results are presented in Tab.~\ref{Tab1}.
The value of the fit parameter $\nu = \left(0.35409\pm 2.43173\times 10^{-5}\right)\frac{\Delta}{2\pi}$
corresponding to the first peak of the power spectrum in Fig~\ref{fig:fig2}(b). 
The remaining higher frequency peaks are located at about $n\nu, n=2\ldots7$.
Tab.~\ref{Tab1} shows that the amplitudes $A_n$ satisfy $|A_2|> |A_1| > |A_3| > \ldots > |A_7|$ while 
the phases $\phi_n$ alternate in their sign.
In Fig.~\ref{fig:fig2}(a) we compare the optimal control field selected by the conjugate
gradient method with the model defined by Eq.~(\ref{eq:fit}).

\begin{figure}
\includegraphics[width=6cm,angle=-90]{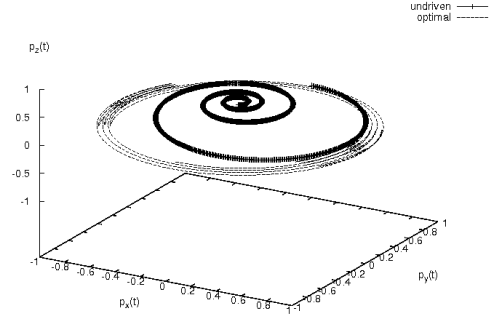}
\caption{Implementation of the Z-gate with 
${\bf p}_I=(0,1,0)^T$ and ${\bf p}_d=(0,-1,0)^T$.
The 3-dimensional plot of the the Bloch vector
${\bf p}=\left(p_x, p_y, p_z\right)^T$
for undriven case and for 
driven by the optimal control field
obtained by the conjugate gradient method
is presented. Parameters as in Fig.~\ref{fig:fig1}.}
\label{fig:bfig3}
\end{figure}

\begin{figure}
\includegraphics[width=7cm,angle=-90]{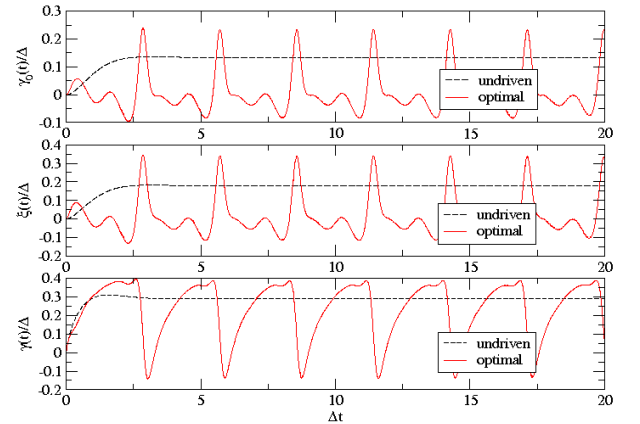}
\caption{Implementation of the Z-gate with 
${\bf p}_I=(0,1,0)^T$ and ${\bf p}_d=(0,-1,0)^T$.
Depicted are the decay rate $\gamma$,
the Lamb shift $\xi$ and the inhomogeneous term $\gamma_0$ 
as a function of time for undriven case and driven by the optimal control 
field selected by the conjugate
gradient method. (a), (b) and (c) show, respectively, the results for $\gamma_0$, $\xi$
and $\gamma$. Parameters as in Fig.~\ref{fig:fig1}.}
\label{fig:bfig4}
\end{figure}

We also studied flipping from state ${\bf p}_I=(0,1,0)^T$ to ${\bf p}_d=(0,-1,0)^T$.
The results are presented in Figs.~\ref{fig:bfig1}-\ref{fig:bfig4}.  
The same picture emerges. The optimized field immediately drives
the system into state ${\bf p}=(1,0,0)^T$,  performs switching between the decoherence-free states 
${\bf p}=(1,0,0)^T$ and ${\bf p}=(-1,0,0)^T$, and finally 
transfers it into the target state ${\bf p}=(0,-1,0)^T$.
Actually, Fig~\ref{fig:bfig2} displays the time evolution of the selected optimal field
and its power spectrum. It is seen in Fig.~\ref{fig:bfig2}(a) 
that the optimal control field starts out positive value
to transfer the system from ${\bf p}_I=(0,1,0)^T$ to ${\bf p}=(1,0,0)^T$
and goes negative value (approximately $-\varepsilon_{0z}$) to trap 
the system in this state. The switching is performed by a positive pulse 
which is optimized to rotate the Bloch vector into state ${\bf p}=(-1,0,0)^T$.
Then the field goes negative again to trap the system in this state.
After, performing several abrupt switch operations between 
the free-decohrence states ${\bf p}=(1,0,0)^T$, ${\bf p}=(-1,0,0)^T$,
the control field value at the final time $t_F$
is positive in order to transfer the system into 
the target state ${\bf p}=(0,-1,0)^T$.
Contrary to the first example, 
the configurations ${\bf p}=(0,p_y,0)^T$ for $-1\leq p_y\leq +1$ are 
not stable under external driving by a negative static control field ${\varepsilon_z}(t)=-\varepsilon_{0z}$.
Thereby, the control optimum field value is positive at the beginning and also 
at end of the time evolution interval $[0,t_F]$ allowing the transfer of the system from
${\bf p}=(0,1,0)^T$ to ${\bf p}=(1,0,0)^T$ at the initial time and from ${\bf p}=(1,0,0)^T$
to the target state ${\bf p}=(0,-1,0)^T$ at a final time as it is illustrated in 
Figs.~\ref{fig:bfig1} and \ref{fig:bfig2}(a).
Fig~\ref{fig:bfig2}(b) shows that the power spectrum displays  
seven pronounced peaks at equidistant frequencies
similar to the first example. The fitting model defined by Eq.~(\ref{eq:fit})
can be used for this case too which it is not shown here for brevity.


For the two examples of implementing a quantum Z-gate,
The conjugate gradient method selects a "multi--component low--frequency".
This aspect of the optimal control field is remarkable.  
Firstly, the optimum field is a superposition of harmonics.  This allows one to 
identify rather small number of optimization parameters for a direct optimization scheme, such as a 
genetic code. Secondly, all essential frequency components lie below the the Ohmic cut--off frequency $\omega_c=20\Delta$.
Hence, we have shown that there are optimized solutions for decoupling system from environment 
at lower frequence than required in the ``bang--bang" approach.  

The presented solutions was obtained by starting from control field zero and the optimization 
algorithm obtained, within the specified cost functional, a solution which performs 7 switching operations.   
In principle, one switching operation would be sufficient. 
Due to the possibility to dynamically create stable intermediate states one is in a 
similar position as with transferring an electron in an isolated two level system. 
In the latter case, increasing the intensity of a resonant harmonic 
light field induces an increasing number of Rabi flip operations.

\subsection{Trapping}

In the following two examples we study the control of the z--component of the Bloch vector, physically, 
corresponding to the spin direction or relative population of "up" and "down" states.    
First we consider trapping of the system in the excited state ${\bf p}_d=(0,0,1)$. 
$p_x=p_y=0$ at the initial time ensures 
that the Bloch vector has vanishing x-- and y--component in the future, 
regardless of the control field applied.
The problem becomes one--dimensional in the Bloch--vector space. 
The chosen cost functional is
\be
\label{eq:J_Trap}
J =\frac{1}{2t_F}\int_0^{t_F}dt\,\Vert {\bf p}(t) -{\bf p}_d\Vert^2
\ee
In this case, the running cost functional follows as 
${\mathcal L}\left({\bf p}(t),{\mbox{\boldmath $\varepsilon$}}(t)\right)=\frac{1}{2t_F}\Vert {\bf p}(t) -{\bf p}_d\Vert^2$
and the final cost functional $\Phi\left\lbrack{\bf p}(t_F)\right\rbrack$ is equal to zero.
For the isolated two--level system there are several known ways of trapping a two--level system by an external control 
with $\sigma_z$--coupling.  One can make the 
trapping state to the ground state of the system or one can apply a monochromatic high--frequency field with matched 
intensity to induce dynamic localization~\cite{Grossmann}.
These strategies can be generalized and be applied to 
the dissipative two--level system~\cite{Grifoni,Stockburger}.
Both strategies have in common that one tries to  find a control field which makes the
trapping--state to an element of the decoherence--free subspace of the driven system. 
Following the first strategy, a static control field can be found to make the state ${\bf p}=(0,0,p_z)^T$ for $-1\leq p_z\leq +1$
to the thermal equilibrium ground state of the driven system for given finite temperature.   
Alternatively, a high--frequency field can be used to dynamically decouple the open quantum system from the bath.
In the ``bang--bang" method mentioned in the introduction, this is achieved with a control field whose frequency is (much) higher than
the maximum frequency of the bath~\cite{Viola_1,Viola_2,Stigmunt_1,Jirari_2}.
In the present model this is the phonon cut--off frequency $\omega_c$.   
Here we will show that a dynamic decoupling can be achieved by a field whose characteristic 
angular frequencies lie below $\omega_c$.  

\begin{figure}
\includegraphics[width=7cm,angle=-90]{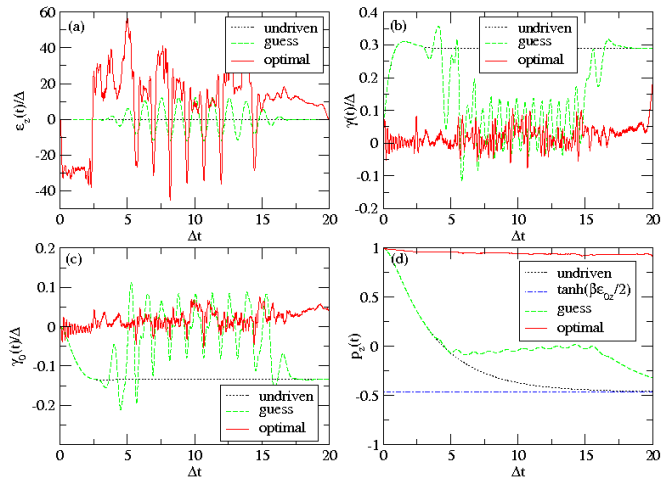}
\caption{Trapping the system in an unstable quantum state,
{\it i.e}, ${\bf p}(t)=(1,0,0)^T$ for all $t \in [0,t_F]$.
Depicted are the control field $\varepsilon_{z}$, the decay rate $\gamma$,
the inhomogeneous term $\gamma_0$ and the relative population
$p_z$ as a function of time 
for three cases of undriven, driven by an harmonic field
with a Gaussian shape (the guessed control field) and 
driven by the optimal control field selected by the conjugate
gradient method.
(a) shows the results for the control field $\varepsilon_{z}$,
(b) and (c) show, respectively, the results for $\gamma$ and $\gamma_0$
while (d) show the results for $p_z$.
The parameters used are $\alpha=0.2$, $\varepsilon_{0z}=-\Delta$,
$\omega_c=20\Delta$ and $k_BT=\beta^{-1}=\hbar\Delta$.
The final time is set as $t_F= 20/\Delta$ and the chosen time step is $10^{-2}/\Delta$
corresponding to $M=2\times 10^3$ mesh points, {\it i.e},
the dimension of the optimal control problem.}
\label{fig:fig6}
\end{figure}

\begin{figure}
\includegraphics[width=6cm,angle=-90]{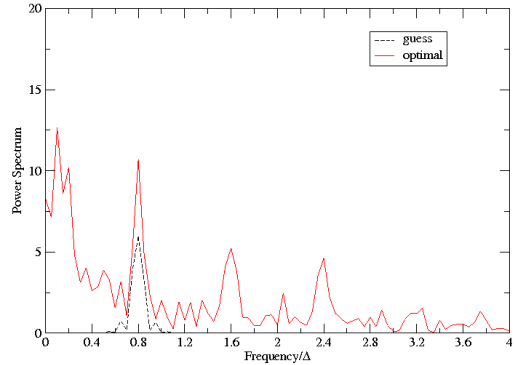}
\caption{Trapping the system in an unstable quantum state
{\it i.e} ${\bf p}(t)=(1,0,0)^T$ for all $t \in [0,t_F]$.
Depicted is the comparison of the power spectrum of the optimal control field selected by the conjugate
gradient method with the power spectrum of the guessed control field.
Parameters as in Fig.~\ref{fig:fig6}.}
\label{fig:fig7}
\end{figure}

In the present model, an oscillating  
control field leads to a rapidly oscillating integrand for $\gamma(t)$ and $\gamma_0(t)$ leading 
to small values for these two functions.  
Fig.~\ref{fig:fig6}(d) shows the time evolution of  $p_z$.   The dotted line shows the 
free evolution of the system into its thermal--equilibrium ground state within a time of about $20/\Delta$. 
Starting from a guess for  the control field in form of a Gaussian pulse, an optimized solution is obtained via 
the conjugate gradient method which stabilizes the system in state ${\bf p}=(0,0,1)^T$ rather well.     
Comparing, the initial guess to the selected optimal control field 
one sees in Fig.~\ref{fig:fig6}(a) that the oscillations of the Gaussian 
pulse get picked up and are amplified.  In regions were the Gaussian factor 
suppressed the field the selected optimal field is less structured. 
Fig.~\ref{fig:fig7} shows the power spectrum of original guess and the selected optimal control field.
The main peak from the original guess gets amplified and higher harmonics of the central 
frequency of the original  guess are used to fine tune the control field.  
The selected field still shows clear features of the original guess. 
This is quite typical for solutions obtained 
within the conjugate gradient method when more than one solution exists.  
Figs.~\ref{fig:fig6}(b) and (c) show that state trapping 
is indeed caused by dynamic decoupling in this case.   
$\gamma(t)$ and $\gamma_0(t)$ show high frequency oscillations of small amplitude about zero.
 
\begin{figure}
\includegraphics[width=6cm,angle=-90]{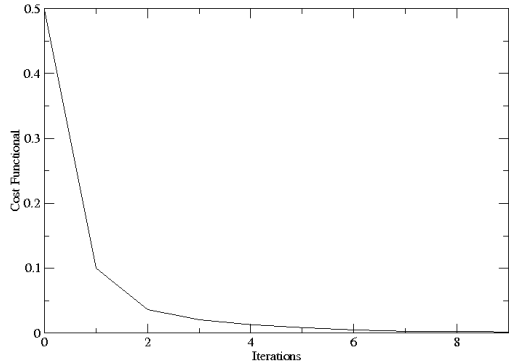}
\caption{Trapping the system in an unstable quantum state
{\it i.e} ${\bf p}(t)=(1,0,0)^T$ for all $t \in [0,t_F]$.
Shown is the cost functional vs. the number of iterations.
Parameters as in Fig.~\ref{fig:fig6}.}
\label{fig:fig8}
\end{figure}

To address the issue of convergence of the numerical procedure we show the cost functional 
(\ref{eq:J_Trap}) versus the number of iterations in Fig.~\ref{fig:fig8}.  
It can be seen that, starting from a mediocre guess, 
convergence is reached typically within  10 iterations.   
Moreover, convergence is strictly monotonic. 
Compared to direct approaches, such as a genetic code, this method requires significantly lower number of 
computations of the cost functional and significantly less computation time.  Thus the investment 
of setting up the optimization scheme by introducing the co--state and the extra task of backward integration 
for the latter pays off in the end.  Moreover,  the present method makes feasible the selection of "arbitrary" optimized 
control fields, {\i. e.} an optimization of the control field at every mesh point in time.  Due to the large 
number of mesh points this would make a direct optimization approach computationally highly expensive. 
There are, however, non--linear programming approaches which may fair well for the present system
~\cite{Betts}.

\subsection{Inversion of population}

As a final example we consider the task of driving the system from its thermal equilibrium state 
$(0,0,p_{z})$
into the pure "up" state $(0,0,1)$ and subsequent trapping in this state.   
The general cost functional given by Eq.~(\ref{eq:cost_Fuctional}) is adapted to the present task 
by setting 
\be
\Phi\left\lbrack{\bf p}(t_F)\right\rbrack = 
\frac{w_1}{2}\Vert p_z(t_f)-p_{zd}(t_f)\Vert^2
\ee
and 
\be
{\mathcal L}\left({\bf p}(t),{\mbox{\boldmath $\varepsilon$}}(t)\right)=
\frac{w_2}{2t_f}\Vert p_z(t)-p_{zd}(t)\Vert^2\
+\frac{1}{2}
s(t)\varepsilon^2(t)~~.
\ee
$s(t)$, $w_1$ and $w_2$ with $w_1+w_2=1$ are real-valued weight factors to specify driving ($w_1=1$) and trapping ($w_2=1$). 
One can use the latter two to shift significance between driving into a target state and driving the system 
along a specified trajectory $p_{zd}(t)$. The function $s(t)$ may be used to tailor the control pulse shape.  
In case of certain linear control problems the third term is necessary to make the problem regular~\cite{Krotov}.
$-1\leq p_{zd}(t)\leq +1$ defines the ``desired" trajectory of the system. 
For the present discussion we set $p_{zd}(t)=1$.

\begin{figure}
\includegraphics[width=7cm,angle=-90]{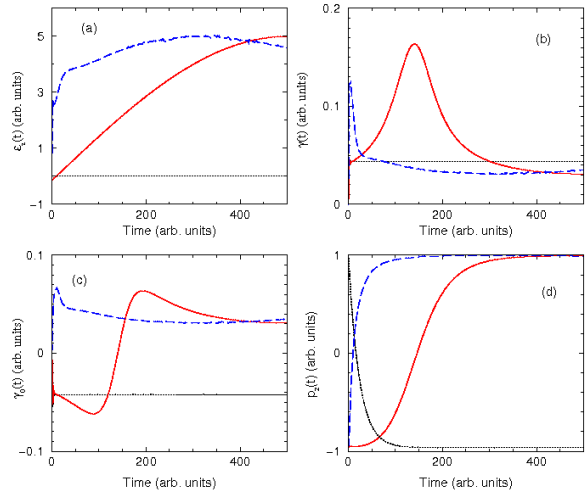}
\caption{Control of relative population $p_z(t)$: 
(a) shows the selected control field,
(b) shows $\gamma(t)$, 
(c) shows $\gamma_0(t)$ and (d) shows the corresponding time evolution.   
The dotted lines are for $\varepsilon(t)$ =0 (undriven case), the solid line is for 
transfer from  $(0,0,-0.965)$ to $(0,0,1)$, 
and the dashed lines are for transfer to and trapping in $(0,0,1)$.
The parameters used are $\alpha=0.25$, $\varepsilon_{0z}=-2$, $\Delta=0.5$,
$\omega_c=10$, $t_F=500$, and temperature $\beta^{-1}=0.5$.}    
\label{Z-M1}
\end{figure}

Numerical results are shown in Fig.~(\ref{Z-M1}).
Let us first look at the undriven case for an initial state $(0,0,1)$, displayed by the dotted lines.  It is seen 
in Fig.~(\ref{Z-M1})(d) that, on the time scale considered, there is
rapid thermalization of the system into the equilibrium state at about $(0,0,-0.96)$.  
Except for oscillations at very short times, $\gamma(t)$ and $\gamma_0(t)$ are essentially constant in time.  

For the driven case we consider two situations.  In the first we wish to prepare the system 
in the target state $(0, 0,1)$  at $t_F=500$ when the system initially is in the thermal equilibrium state 
$(0,0,-0.964)$.  We set $w_1=1$ and $ w_2=0$.  Since the intrinsic 
time scale is faster than the target time there exist many solutions to achieve the task.  Here we
choose an initial guess in form of a harmonic field of low frequency (adiabatic solution) and optimize 
this guess subsequently with the conjugate gradient method using 300 mesh points for the control field.
We use $s(t)$ to suppress the control 
field at times around zero, as well as high intensities.   Results for an optimal solution are shown by the solid 
lines in Fig.~(\ref{Z-M1}).  The selected optimal control fulfils the conditions imposed and leads to 
a gradual transition into the target state.  In this particular case, the qubit--environment coupling has 
effectively been reduced over the undriven case.  

The second case considered is driving the system from its thermal equilibrium state into the target 
state $(0,0,1)$ as fast as possible and subsequently trap it there.   In the cost functional we set 
$w_1=0$ and $ w_2=1$.   Again a low--frequency harmonic field is 
selected for the initial guess and $s(t)$ is used to tailor the selected control 
field at times around zero, as well to limit its intensity.  The results are shown 
by the dashed lines in Fig.~(\ref{Z-M1}).  The selected control field rises sharply from about zero
to, essentially, a plateau.  $\gamma(t)$ and $\gamma_0(t)$ vary significantly only in the time during the transfer.  
Although high fields are suppresses around time zero, the selected optimal control 
field manages a more rapid transfer into the new equilibrium case (dashed line in Fig.~(\ref{Z-M1})(d)) 
than the undriven case (dotted line in Fig.~(\ref{Z-M1})(d)).  
Hence, we show that we have been able to significantly increase the effective interaction strength.

\section{SUMMARY AND CONCLUSIONS\label{VI}}

In this work we have presented dynamic control of open quantum systems as an optimization problem.  
The Bloch--Redfield approach was used to derive Markovian kinetic equations of a driven open quantum system 
whereby the external control was treated non--perturbatively. 
This approach leads to a Redfield tensor which accounts for a coupling between system and bath 
which contains a causal dependence upon the external control field.  Indeed, the present approach is equivalent 
to the time--convolutionless projection operator method within second order in the system--environment coupling~\cite{Breuer}.
This control--field dependence of the effective system interaction allows steering of the open quantum system
and its coupling to its environment beyond what is feasible within a semiclassical treatment 
of the environment in which interference between the system--control--field interaction and 
the system--environment interaction is neglected~\cite{Jirari}. 

This approach was applied to the spin--boson model in the strong electron--boson coupling limit.  
Using the polaron transformation, the kinetic equations for the Bloch vector were derived and analysed.   
They feature an effective coupling in the spin system which is renormalized by the spin--phonon interaction and displays 
a  causal dependence (non--local in time)  on the control field.  Analytic results for Lamb shift and decay time are presented for the zero 
temperature limit in the absence of the control.
It is shown for several examples that both the stationary states of the driven open quantum system and the rates at which they are reached can be controlled to a large degree. 

Steering of the open quantum system is formulated and solved as an optimization problem via Pontryagin's minimum principle which is based on 
the introduction of Lagrangean multipliers in form of a  co--state (adjoint state).  The set of optimality
conditions is solved iteratively using a conjugate gradient method.  Numerical examples show that it leads to a monotonic 
improvement in the cost functional.

Several physical situations have been investigated numerically to demonstrate quantum--interference--based optimal control of open quantum system. The studies of a $\pi$ rotation of the Bloch vector in the x--y plane and trapping along the z--axis have shown 
that optimal control fields of moderate frequency (as compared to the phonon cut--off frequency) can be selected 
which significantly extend the lifetime of purity and, hence, improve the chance of successful completion of 
an error free quantum operation or the storage of a dissipative system in a fixed quantum state.  
The analysis of driving and subsequent trapping into a quantum state, which for the undriven system is highly unstable, at the example of population transfer has shown that this task can be achieved by slowly varying fields
for the present model.   Moreover, the rate of transfer can be varied within limits set by the maximally 
obtainable effective coupling strength of the open quantum system.   The latter is determined by the system, the environment, and the system--environment coupling.  

This analysis has also shown that the inverse problem of identification of optimal control fields 
in general has a large number of optimal solutions.   Within the conjugate gradient method the selected optimal solutions usually show a remnance of the initial guess.   The number of optimal solutions may be reduced by 
additional constraints which may be used to select experimentally feasible solutions, such as fields with a gradual rise time, rather than abruptly turned on fields.  In some cases, quite different fields can produce 
near equal results.   For example,  trapping in a quantum state may be obtained by applying a static control field 
which makes the trapping state to its (approximate) new ground state.  In this case a decoupling of the system--environment interaction is not necessary or even desirable.   It is in fact the system--bath coupling which drives the system  system into its new equilibrium state.   
As an important result this study has shown that state--specific optimal control can be achieved by time--dependent 
fields whose characteristic frequencies lie below the maximum characteristic bath frequency.  
Alternatively, high--frequency high--amplitude ``bang--bang" control  fields, reminiscent of the effect of dynamic localization, may induce dynamic decoupling by making the effective coupling strength small. 

Optimization of a dissipative quantum gate poses a more complicated problem than the one addressed 
here since optimization should occur independent of the input state~\cite{Cirac,Thorwart}.
Moreover, the output state (target state), in general, depends on the input state.  
We find that an optimal control field 
critically depends on the input state.   A bang--bang solution (which is probably difficult to implement in experiment) 
can be envisioned whereby a high--frequency high--intensity field is applied to suppress the effective system--environment coupling.  
However, such a field usually also has a direct coupling channel to the 
system which may cause problems when the control field is not perfectly suitable for the input state.  
Whether the present approach which is based on specific trajectories can be extended to optimize quantum gates 
by some averaging procedure or whether an optimization should directly be aimed at the superoperator 
responsible for the time--evolution will require further investigation.

\begin{acknowledgments}
We wish to acknowledge financial support of this work by FWF, project 
number P16317-N08.
\end{acknowledgments}

\appendix

\section{SCHWINGER AND FEYNMAN REPRESENTATIONS\label{APP_A}}

The Schwinger and Feynman representations~\cite{Peskin}
will play an important role in the determination of the decay  
rate and the Lamb shift (see below).

The Schwinger representation involves the Euler Gamma function defined by
\be
\label{eq:Euler}
{\bf\Gamma(\nu)}=\int_0^{\infty}\,dt\,e^{-t}\,t^{\nu-1},\qquad{\rm Re}\,\nu > 0,
\ee
Making the variable change $Du=t$ in the definition of the Euler gamma function 
(\ref{eq:Euler}), leads to
\be
\label{eq:Schwinger_I}
\frac{1}{D^\nu} =\frac{1}{{\bf\Gamma(\nu)}}
\int_0^{\infty}\,du\,u^{\nu-1}e^{-uD},\quad\,{\rm Re}\,\nu > 0\quad{\rm Re}\,D > 0.
\ee
The identity (\ref{eq:Schwinger_I})
allows to write the denominators $D$ of the propagator in form of an integral
on the Schwinger parameter $u$.

On the other hand the Feynman representation~\cite{Peskin} 
introduces a parameter $x$ (Feynman parameter) to squeeze the denominator factors
into a single polynomial form
\be
\label{eq:Feynman_I}
\frac{1}{A\centerdot B}=\int_0^1\,dx\,\left\lbrack Ax + B(1-x)\right\rbrack^{-2}\,.
\ee
%
\begin{figure}
\includegraphics[width=7cm,angle=-90]{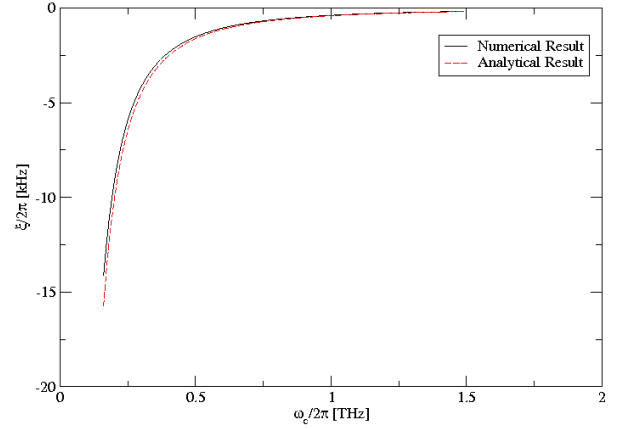}
\caption{The Lamb shift for the undriven case and zero temperature
as a function of the Ohmic cutoff frequency $\omega_c$.
The test of our analytical result, Eq.~(\ref{eq:C_3}),
by comparison of a direct numerical integration of 
Eq.~(\ref{eq:xi}) is shown. The parameters used are the coupling $\alpha = 1.2$,
the tunnel amplitude $\frac{\Delta}{2\pi}=635\,\mbox{MHz}$
and the energy bias $\frac{\varepsilon_{0z}}{2\pi}= 560\,\mbox{MHz}$.}
\label{fig:XI}
\end{figure}

\section{DERIVATION OF THE DECAY AT ZERO TEMPERATURE\label{APP_B}}

Here we derive an analytical expression of the decay rates 
\bea
\gamma(\varepsilon_{0z}) &=&
\Delta^2\int_0^{\infty}\,d\tau\,e^{-Q_2(\tau)}\cos[Q_1(\tau)]\cos[\varepsilon_{0z}\tau]\nonumber\\
&=& \gamma_f(\varepsilon_{0z})+\gamma_b(\varepsilon_{0z})\,,
\eea
with
\be
\label{eq:gamma_f}
\gamma_f(\varepsilon_{0z})=
\frac{\Delta^2}{2}
{\rm Re}\int_0^{\infty}\,
d\tau\,
e^{-Q_2(\tau)}
e^{iQ_1(\tau)}
e^{-i\varepsilon_{0z}\tau}\,,
\ee
and 
\be
\gamma_b(\varepsilon_{0z})=
\frac{\Delta^2}{2}
{\rm Re}\int_0^{\infty}\,
d\tau\,
e^{-Q_2(\tau)}
e^{iQ_1(\tau)}
e^{i\varepsilon_{0z}\tau}
\ee
are, respectively, the forward and the backward decay rates.
Note that $\gamma_b(\varepsilon_{0z}) = \gamma_f(-\varepsilon_{0z})$.
Substituting Eqs.~(\ref{eq:Q_1}) and (\ref{eq:Q_2}) into Eq.~(\ref{eq:gamma_f}),
we obtain 
\be
\label{eq:bis_gamma_f}
\gamma_f(\varepsilon_{0z})=
\frac{\Delta^2}{2}
{\rm Re}\int_0^{\infty}
\frac{e^{-i\varepsilon_{0z}\tau}}{{(1-i\omega_c\tau)}^{2\alpha}}\,d\tau\,.
\ee
Now with the help of the Schwinger identity (\ref{eq:Schwinger_I}),
Eq.~(\ref{eq:bis_gamma_f}) can be written as 
\be
\label{eq:bbis_gamma_f}
\gamma_f(\varepsilon_{0z})=
\frac{\Delta^2}{2{\bf\Gamma}(2\alpha)}
{\rm Re}\int_0^{\infty}du\,u^{2\alpha-1}\,e^{-u}\,
\int_0^{\infty}d\tau
e^{-i(\varepsilon_{0z}-u\omega_c)\tau}\,.
\ee
Using the fact that
\be
\label{eq:dirac}
\int_0^{\infty}d\tau
e^{-i(\varepsilon_{0z}-u~\omega_c)\tau} =
\pi\,\delta(\varepsilon_{0z}-u~\omega_c) - 
i~\mbox{PP}\left(\frac{1}{\varepsilon_{0z}-u\omega_c}\right)\,,
\ee
where the first term is the Dirac distribution
and the second term $\mbox{PP}$ denotes the Cauchy principal
part of the integral 
$
\int_0^{\infty}d\tau/\left(\varepsilon_{0z}-u\,\omega_c\right)
$ and introducing the Heaviside distribution $\theta(u)$ to extend
the bounds of integration from $-\infty$ to $+\infty$,
Eq.~(\ref{eq:bbis_gamma_f}) becomes  
\be
\label{eq:bbbis_gamma_f}
\gamma_f(\varepsilon_{0z})=
\frac{\pi\Delta^2}{2{\bf\Gamma}(2\alpha)\,\omega_c}
\int^{+\infty}_{-\infty}du\,u^{2\alpha-1}\,e^{-u}\,\theta(u)\,
\delta\left(u-\frac{\varepsilon_{0z}}{\omega_c}\right)\,.
\ee
Evaluating the convolution product (\ref{eq:bbbis_gamma_f}),
one ends up with the following formula:
\be
\gamma_f(\varepsilon_{0z})=\frac{\pi\Delta^2}{2{\bf\Gamma}(2\alpha)}\left(\frac{1}{\omega_c}\right)^{2\alpha}\,
\varepsilon^{2\alpha-1}_{0z}\,e^{-\left(\varepsilon_{0z}/\omega_c\right)}\,
\theta\left(\varepsilon_{0z}/\omega_c\right)\,.
\ee
The Heaviside distribution (or step function), 
insures that at zero temperature absorption of energy
from the environment is not possible.
The final result for the decay rate is then for $\alpha > \frac{1}{2}$,
\begin{subequations}
\bea
\gamma(\varepsilon_{0z} > 0)&=&\frac{\pi\Delta^2}{2{\bf\Gamma}(2\alpha)}\left(\frac{1}{\omega_c}\right)^{2\alpha}
\varepsilon^{2\alpha-1}_{0z}\,e^{-\left(\varepsilon_{0z}/\omega_c\right)}\,,\\
\gamma(\varepsilon_{0z} < 0)&=&\frac{\pi\Delta^2}{2{\bf\Gamma}(2\alpha)}\left(\frac{1}{\omega_c}\right)^{2\alpha}
\left(-\varepsilon_{0z}\right)^{2\alpha-1}\,e^{\left(\varepsilon_{0z}/\omega_c\right)}.\nonumber\\
&&
\eea
\end{subequations}
For the inhomogeneous term
\be
\gamma_{0}(\varepsilon_{0z})=
\Delta^2\int_0^{\infty}\,d\tau\,e^{-Q_2(\tau)}\sin[Q_1(\tau)]\sin[\varepsilon_{0z}\tau]\,,
\ee
similar calculation of the decay rate leads for $\alpha > 1/2$ to
\begin{subequations}
\bea
\gamma_0(\varepsilon_{0z} > 0)&=&\frac{\pi\Delta^2}{2{\bf\Gamma}(2\alpha)}\left(\frac{1}{\omega_c}\right)^{2\alpha}
\varepsilon^{2\alpha-1}_{0z}\,e^{-\left(\varepsilon_{0z}/\omega_c\right)}\,,\\
\gamma_0(\varepsilon_{0z} < 0)&=&-\frac{\pi\Delta^2}{2{\bf\Gamma}(2\alpha)}\left(\frac{1}{\omega_c}\right)^{2\alpha}
\left(-\varepsilon_{0z}\right)^{2\alpha-1}\,e^{\left(\varepsilon_{0z}/\omega_c\right)}.\nonumber\\
&&
\eea
\end{subequations}
At zero temperature, the detailed balance condition takes the following form
\be
\frac{\gamma_0(\varepsilon_{0z})}{\gamma(\varepsilon_{0z})}=
\left\{
\begin{array}{cc}
1 &\quad\mbox{if}\quad\varepsilon_{0z} > 0\\
-1 &\quad\mbox{if}\quad\varepsilon_{0z} < 0
\end{array}
\right.
\ee
\begin{figure}
\includegraphics[width=7cm,angle=-90]{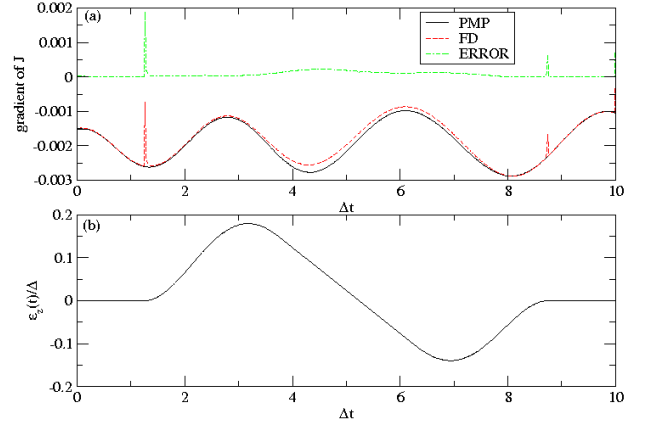}
\caption{Implementation of the Z-gate with 
${\bf p}_I=(1,0,0)^T$ and ${\bf p}_F=(-1,0,0)^T$.
The upper panel (b) shows the numeric comparison between Pontryagin's minimum principle (PMP) 
and the finite difference (FD) approximation for computing the gradient of the final cost functional 
in Eq.~(\ref{eq:Z_J}) evaluated at the control field displayed in the lower panel (b).
The parameters used are $\alpha=0.2$, $\varepsilon_{0z}=\Delta$,
$\omega_c=20\Delta$, $k_BT=\beta^-1=\hbar\Delta$.
The final time is set as $t_F= 10/\Delta$ and the chosen time step is $5\times10^{-2}\Delta$
corresponding to $M=2\times 10^2$ mesh points, {\it i.e},
the dimension of the optimal control problem.}
\label{fig:grad}
\end{figure}

\section{DERIVATION OF THE LAMB SHIFT AT ZERO TEMPERATURE\label{APP_C}}

Let us now compute the Lamb shift given by
\bea
\label{eq:xi}
&&\xi(\varepsilon_{0z})=
\Delta^2\int_0^{\infty}d\tau\,e^{-Q_2(\tau)}\cos[Q_1(\tau)]\sin[\varepsilon_{0z}\tau]\nonumber\\
&&
=\frac{\Delta^2}{2}
{\rm Im}\int_0^{\infty}
d\tau
e^{-Q_2(\tau)}
\left\{
e^{iQ_1(\tau)}
e^{i\varepsilon_{0z}\tau}
+ 
e^{-iQ_1(\tau)}
e^{i\varepsilon_{0z}\tau}
\right\}\,.
\nonumber\\
&&
\eea
Substituting Eqs.~(\ref{eq:Q_1}) and (\ref{eq:Q_2}) into 
Eq.~(\ref{eq:xi}) and using (\ref{eq:dirac}), we obtain  
\be
\label{eq:bxi}
\xi(\varepsilon_{0z})=-\frac{\Delta^2}{{\bf\Gamma}(2\alpha)}
\frac{\varepsilon_{0z}}{\omega^2_c}\int_0^{\infty}\,du
\frac{u^{2\alpha-1}\,e^{-u}}{u^2-\left(\frac{\varepsilon_{0z}}{\omega_c}\right)^2}\,,
\ee
where $u$ is the dimensionless Schwinger parameter.
The last integral can not be computed using the residues theorem
since it is singular at $u=\pm\frac{\varepsilon_{0z}}{\omega_c}$
and at $u=0$ when $\alpha < 1/2$.

Nevertheless the application of the Feynman identity (\ref{eq:Feynman_I})
to Eq.~(\ref{eq:bxi}), leads to
\be
\label{eq:bbxi}
\xi(\varepsilon_{0z}) = -\frac{\Delta^2}{{\bf\Gamma}(2\alpha)}
\frac{\varepsilon_{0z}}{\omega^2_c}\int_0^1\,dx\,{\mathcal I}
\left(x,\alpha,\frac{\varepsilon_{0z}}{\omega_c}\right)\,,
\ee 
with
\be
{\mathcal I}
\left(x,\alpha,\frac{\varepsilon_{0z}}{\omega_c}\right)
=\int_0^{\infty}\,du
\frac{u^{2\alpha-1}\,e^{-u}}{\left(u+(2x-1)\frac{\varepsilon_{0z}}{\omega_c}\right)^2}
\ee
which after the change variable $v=u+(2x-1)\frac{\varepsilon_{0z}}{\omega_c}$
is transformed to 
\bea
{\mathcal I}\left(x,\alpha,\frac{\varepsilon_{0z}}{\omega_c}\right)&=&
\int_{(2x-1)\frac{\varepsilon_{0z}}{\omega_c}}^{\infty}\,dv
\left(v-(2x-1)\frac{\varepsilon_{0z}}{\omega_c}\right)^{2\alpha-1}\nonumber\\
&&\quad\quad\quad\times v^{-2}e^{-v}\,e^{(2x-1)\frac{\varepsilon_{0z}}{\omega_c}}\,.
\eea
Now, the approximation $\frac{\varepsilon_{0z}}{\omega_c}\ll 1$
leads to
\bea
{\mathcal I}\left(x,\alpha,\frac{\varepsilon_{0z}}{\omega_c}\right)&=&
e^{(2x-1)\frac{\varepsilon_{0z}}{\omega_c}}\,
\int_0^{\infty}\,dv\,v^{2\alpha-3}\,e^{-v}
\nonumber\\
&=&
\label{eq:I_x}
e^{(2x-1)\frac{\varepsilon_{0z}}{\omega_c}}\,{\bf\Gamma}(2\alpha-2)\,.
\eea
%
%
Using Eq.~(\ref{eq:I_x}), we get from (\ref{eq:bbxi}) for $\alpha > 1$
\be
\label{eq:C_3}
\xi(\varepsilon_{0z})=-\frac{\Delta^2}{\omega_c^2}\,
\varepsilon_{0z}\,
\frac{\sinh\left(\varepsilon_{0z}/\omega_c\right)}{\left(\varepsilon_{0z}/\omega_c\right)}\,
\frac{{\bf\Gamma}(2\alpha-2)}{{\bf\Gamma}(2\alpha)}\,.
\ee
Combining again the Schwinger identity (\ref{eq:Schwinger_I})
with Eq.~(\ref{eq:I_x}) and after some algebra we obtain for 
$\frac{1}{2}<\alpha < 1$
\bea
\label{eq:bbbxi}
&&\xi(\varepsilon_{0z}) = -\frac{\Delta^2}{2}
\frac{\varepsilon_{0z}}{\omega^2_c}\,
\left({\frac{\varepsilon_{0z}}{\omega_c}}\right)^{2\alpha-2}
{\bf\Gamma}(2-2\alpha)\,\frac{1}{2\alpha-1}\nonumber\\
&&\quad\quad\quad\quad\times
\lim_{\eta\to 0}\left(1-e^{(2\alpha-1)\log(2\eta-1)}\right)\,.
\eea
A such limit in the last equation does not exist. 

I summary, our prediction for the renormalization of the energy bias due to the Lamb shift
at zero temperature; in leading order in $\frac{\varepsilon_{0z}}{\omega_c}$
($\frac{\varepsilon_{0z}}{\omega_c} \ll 1$) and in strong coupling regime 
($\alpha > 1$), is the following:
\be
\tilde{\varepsilon}_{0z}=
\varepsilon_{0z}\times
\left\lbrack
1
-\frac{\Delta^2}{\omega_c^2}\,
\frac{\sinh\left(\varepsilon_{0z}/\omega_c\right)}{\left(\varepsilon_{0z}/\omega_c\right)}\,
\frac{{\bf\Gamma}(2\alpha-2)}{{\bf\Gamma}(2\alpha)}
\right\rbrack\,.
\ee

The agreement of the analytical expression for the Lamb shift, Eq.~(\ref{eq:C_3}), 
with the numerical integration of Eq.~(\ref{eq:xi}) by Gauss quadrature is shown 
in Fig.(\ref{fig:XI}).

\section{NUMERICAL TEST OF THE GRADIENT}

In order to test the method of Pontryagin's minimum principle (PMP) 
given by Eq.~(\ref{eq:dj_de_k}) to compute the gradient of the cost functional, 
Eq.~(\ref{eq:Z_J}), we compare it with  
the finite difference approximation (FD). 
Fig.~\ref{fig:grad} shows the result of the gradient for a control field of the form 
$\varepsilon_z(t)=A(t)\cos(\Omega t+\phi)$ with  frequency $\Omega$,
phase $\phi$, and a Gaussian envelope $A(t)$.
We can see in Fig.~\ref{fig:grad} that the error, 
$
\mbox{ERROR}=\left|\nabla^{\rm PMP} J - \nabla^{\rm FD} J\right|,
$
is roughly zero except at the switching times of the control field
where its amplitude is suddenly increased.
The good agreement observed for this case occurs because the frequency $\Omega=10^{-1}\times\Delta$ 
of the control field 
is low and causes a slow variation of the cost functional.
In this case the finite difference 
approximation is numerically stable and gives good results
compared to the adjoint--state method.
In case of high control field, this good agreement is lost because 
the cost functional varies very rapidly and renders 
the finite difference method numerically unstable.


\end{document}